\documentclass[smallextended]{svjour3}     
\usepackage{soul}
\usepackage{booktabs}
\usepackage{tcolorbox}
\usepackage{xcolor, soul}
\usepackage{csquotes}
\usepackage{multirow}
\usepackage[disable]{todonotes}
\usepackage{dirtytalk}
\usepackage{comment}
\tcbuselibrary{many}
\sethlcolor{lightgray}  
\tcbset{
    enhanced,
    breakable,
    attach boxed title to top left={
        xshift=0.5cm,
        yshift=-\tcboxedtitleheight/2},  
    top=4mm,
    coltitle=white,
    beforeafter skip=\baselineskip,
}
\newenvironment{mybox}[1]{%
    \begin{tcolorbox}[title={Transformed #1}]%
    }{
    \end{tcolorbox}
    }
    \newenvironment{mybox1}[1]{%
    \begin{tcolorbox}[title={Improvements implemented~--~#1}]%
    }{
    \end{tcolorbox}
    }
    
        \newenvironment{mybox2}[1]{%
    \begin{tcolorbox}[title={Implications for practice~--~#1}]%
    }{
    \end{tcolorbox}
    }

\begin{document}
\title{Large scale reuse of microservices using DevOps and InnerSource practices - A longitudinal case study}
\subtitle{}

\author{Deepika Badampudi         \and
        Muhammad Usman \and Xingru Chen 
}

\institute{D. Badampudi \at
              Department of Software Engineering \\
              Blekinge Institute of Technology\\
              Karlskrona, Sweden\\
              \email{deepika.badampudi@bth.se}          
           \and
           M. Usman \at 
              Department of Software Engineering \\
              Blekinge Institute of Technology\\
              Karlskrona, Sweden\\
              \email{muhammad.usman@bth.se} 
           \and   
            X. Chen \at 
              Department of Software Engineering \\
              Blekinge Institute of Technology\\
              Karlskrona, Sweden\\
              \email{xingru.chen@bth.se} 
}

\date{Received: date / Accepted: date}

\maketitle

\begin{abstract}
Contemporary practices such as InnerSource and DevOps promote software reuse. This study investigates the implications of using contemporary practices on software reuse. In particular, we investigate the costs, benefits, challenges, and potential improvements in contemporary reuse at Ericsson. 
We performed the study in two phases: a) the initial data collection based on a combination of data collection methods (e.g., interviews, discussions, company portals), and b) a follow-up group discussion after a year to understand the status of the challenges and improvements identified in the first phase.
Our results indicate that developing reusable assets resulted in upfront costs, such as additional effort in ensuring compliance. Furthermore, development with reuse also resulted in additional effort, for example, in integrating and understanding reusable assets. Ericsson perceived the additional effort as an investment resulting in long-term benefits such as improved quality, productivity, customer experience, and way of working.
Ericsson's main challenge was increased pressure on the producers of reusable assets, which was mitigated by scaling the InnerSource adoption. InnerSource success is evident from the increase in the contributions to reusable assets. In addition, Ericsson implemented measures such as automating the compliance check, which enhanced the maturity of reusable assets and resulted in increased reuse.  
\end{abstract}


%
\def\ADPPM{Program manager }
\def\ADPA{ADP Anchor }
\def\ADPCA{ADP Chief architect }
\def\ADPPO{ADP Product owner }
\def\ADPHead{ADP head }
\def\ADPHL{ADP Hub lead }
\def\PlatM{Platform manager }
\def\PlatA{Platform architect }
\def\BAM{Developer }
\def\AppSPM{Strategic product manager }
\def\AppTL{Tech lead }
\def\AppCM{Product manager }

\section{Introduction}\label{intro}
Software reuse is the practice of building systems using existing software assets. Over the decades, we have observed different reuse assets, for example, from small subroutines to large components providing significant functionality \cite{capilla2019opportunities}. In addition, the type of code reuse assets has continuously evolved. Building systems using reusable web services have been popular since the 1990s, which has evolved into microservices reuse more recently \cite{capilla2019opportunities}. A microservice is defined as \say{a cohesive, independent process interacting via messages} \cite{dragoni2017microservices}. 

The important characteristics of microservice architectures are to create modular and scalable solutions that are resilient to architecture erosion and easy to develop and test \cite{capilla2019opportunities}, which further contributes to increased reuse \cite{gouigoux2017monolith}. Software platforms such as docker and Kubernetes are used for building large-scale applications using microservices. Such platforms are used to achieve better reuse ratios and easy deployment \cite{dragoni2017microservices}. In addition, DevOps practices such as continuous integration and delivery promote the faster deployment of services \cite{mauro2015adopting}, which could be beneficial when integrating common microservices. 

Modern practices such as InnerSource promote reuse \cite{capraro2016inner}. InnerSource provides patterns to eliminate reuse barriers, such as the lack of discoverability of available common services. In addition, InnerSource patterns promote a collaborative approach where potential consumers of common services participate in the analysis, design, development, and testing, resulting in better alignment of consumer needs. Companies adopt InnerSource practices to increase reuse rate \cite{stol2014key}. 

The evolution of software practices has influenced how we practice software reuse today. The most noticeable changes are in the reuse assets (components to microservices), and collaborative practices (InnerSource). We refer to the reuse of microservices using platforms such as dockers, DevOps, and InnerSource practices as contemporary reuse. 

While the benefits of individual contemporary practices on reuse were investigated, the implications of large-scale contemporary reuse are not well known. 
In our previous study, we investigated the implications of software reuse in contemporary software engineering practices in a medium-sized company\cite{chen2022reuse}. 

In this study, we investigate the implications of contemporary reuse, i.e., large-scale reuse of microservices using DevOps and InnerSource practices in a cloud-native environment at Ericsson. More specifically, we identify contemporary reuse's costs, benefits, challenges, and improvement suggestions. We also revisit the challenges after one year of identifying the challenges and improvement suggestions to understand what improvements were implemented and which challenges were mitigated.  The contributions of our study are as follows - 
\begin{itemize}
    \item We provide a detailed description of the contemporary reuse process. 
    \item We identify the costs and benefits of contemporary reuse. 
    \item We identify the challenges and improvement suggestions in contemporary reuse. 
    \item We provide a summary of improvements implemented by Ericsson to mitigate the identified challenges.
\end{itemize}

The remainder of the paper is structured as follows: We describe the related concepts and related work in Section \ref{sec:related work}.  Section \ref{sec:rm} provides the methodology used to answer the research questions. Section \ref{sec:case}, provides a detailed description of the contemporary reuse process followed at Ericsson. In Section \ref{sec:resultsRQ1} and \ref{sec:challenges}, we provide the results on contemporary reuse's costs, benefits, challenges, and improvement suggestions. In addition, we provide details on the follow-up study on contemporary reuse challenges and improvements. We discuss the results of our study in Section \ref{sec:discussion}. Section \ref{sec:conclusion} concludes the paper and proposes future work.

\section{Background and Related Work} \label{sec:related work} \todo[inline]{XCN}

This section introduces concepts related to our study and the related work on software reuse implications. 
\subsection{Studies investigating software reuse implications}
Researchers have investigated software reuse practices for over five decades. In this section, we discuss the related work on the implications of software reuse. 

\textbf{Benefits and costs:} Barros Jutso et al. \cite{barros2018software} conducted a mapping study on software reuse, and they found increased development productivity and better product quality as the top benefits of software reuse. Kruger and Berger \cite{kruger2020empirical} found that software reuse had additional costs in coordinating, designing, and implementing reusable assets. In addition, our previous study \cite{chen2022reuse} identified software reuse benefits in improved learning, increased usability, and optimized resource allocation as well as software reuse costs in reusable assets documentation, additional approvals, and supporting tools, and extra time in learning reusable assets. 

\textbf{Challenges and improvements: }Bauer and Vetro \cite{bauer2016comparing} investigated software reuse challenges in two large-scale companies. They found that the granularity of the reusable assets is challenging for both case companies. One of the case companies experienced challenges in the lack of organizational-wide reuse policy and transparency of the reusable assets. In addition, both companies identified issues related to loss of control of the reusable assets, reduced performance, and not having the possibility to change the reusable assets. Barros-Justo et al. \cite{barros2019exploratory} replicated the study of Bauer and Vetro in a medium-sized company and noted that the reuse challenges in the medium-sized company relate to the discoverability of the reusable assets and adapting them to the project needs. Furthermore, dependency issues, such as versioning dependencies, and ripple effects caused by changing reusable assets are also challenging according to the previous studies \cite{barros2019exploratory,bauer2016comparing}.
The existing literature suggests that the understandability and discoverability of the reusable assets \cite{agresti2011software,bauer2014exploratory,chen2022reuse} and the distribution of the reuse-related tasks among developers \cite{agresti2011software,chen2022reuse} are important for improving software reuse. In addition, improving the reuse measurement to help higher management understand the reuse benefits is also important \cite{chen2022reuse}.

\subsection{Studies investigating microservices as a reusable unit} 
Microservices are small and autonomous services \cite{newman2021building}. Each microservice has its own independent unit of development, deployment, operations, versioning, and scaling \cite{balalaie2016microservices}. 
Furthermore, microservices can be independent when loosely coupled and do not share data unless necessary \cite{newman2019monolith}. Containers and DevOps are used along with microservices for better application development, testing, and deployment \cite{pahl2015containerization,balalaie2016microservices}.

Microservices enable increased reuse and faster delivery; therefore, companies are transitioning from monolith to microservices. Assuncao et al. \cite{carvalho2019analysis,assunccao2022analysis} identified five criteria that help to identify potential candidates to transition to microservices from legacy systems. The identified criteria are coupling, cohesion, feature modularization, network overhead, and reuse. Using the multi-criteria strategy to identify candidates for transitioning to microservices helps increase the reuse opportunities  \cite{assunccao2021multi}. Tizzei and Nery \cite{tizzei2017using} found that microservices and product line architecture support multi-tenant Software as a Service (SaaS) reuse and observed an increase in LOCs reuse by 60\%. Silva et al. \cite{da2022spread} observed that transitioning to microservices and using DevOps improved portability, reuse, and maintainability. Similar findings were made by Shadija et al. \cite{shadija2017microservices} who found more \say{fine-grained} microservices enable greater reuse when the microservices are located in the same container or different containers but on the same host. 

Two studies explicitly mentioned the benefits and challenges of reusing microservices \cite{gouigoux2017monolith,wang2021promises}. Gouigoux and tamzalit \cite{gouigoux2017monolith} identified that microservices increase reuse in different business contexts and decrease time when developing similar applications resulting in financial gain. Independent microservices allow different teams to use different languages. However, Wang et al. \cite{wang2021promises} found that the benefit of using multiple languages limits the reuse opportunity since understanding the microservices in different languages takes a longer time. 
In addition, the authors also found it is challenging to manage the common code shared by multiple microservices.

\subsection{Studies investigating InnerSource as reuse enabler}
InnerSource is adopting open source development practices in in-house software development \cite{aagerfalk2015software}. InnerSource encourages individuals to contribute to other teams to increase the feature implementation, and defect fixing speed \cite{cooper2018adopting}. 

One of the main reasons why companies adopt InnerSource is to increase software reuse \cite{cooper2018adopting,lindman2010open,stol2014key}. InnerSource practices such as building a common space for sharing reusable assets promote reuse as it provides better transparency and avoids rework. The benefit of using common spaces on reuse is observed in companies such as Lucent Technologies \cite{gurbani2006case}, Philips \cite{lindman2010open}, Nokia \cite{lindman2010open}, Hewlett-Packard \cite{melian2008lost}, Ericsson \cite{torkar2011adopting}, and IBM \cite{vitharana2010impact}. A similar benefit was found in a study by Morgan et al. \cite{morgan2021share} investigating the common spaces in Zalando, Philips Healthcare, and PayPal. The common spaces are referred to as software forge, marketplace, portal, or platform by different companies. The shared reusable assets in common spaces could be projects \cite{riehle2009open,lindman2010open}; stand-alone products, frameworks and documentation templates \cite{linaaker2014infrastructure,riehle2016inner}; software products \cite{schreiber2014open}; software modules and items \cite{dinkelacker2002progressive}; and software libraries and components \cite{lindman2010open,morgan2011exploring,ripatti2015internal}. 


In addition to increased transparency, InnerSource practices also facilitate better communication between teams  which results in a better understanding of the reusable assets from the consumers' perspective \cite{gurbani2006case,gurbani2010managing,vitharana2010impact}. Furthermore, Linden \cite{linden2010open} found that InnerSource helps relieve the providers from some consumers' requests since consumers can themselves also contribute to the reusable assets and thus resulting in faster time to market. However, it is not easy to attract InnerSource contributions from developers due to the lack of available capacity \cite{morgan2011exploring} and awareness about the benefits of making InnerSource contributions \cite{wesselius2008bazaar}. In addition, from the producers point of view, accepting contribution may also involve challenges as they need to ensure that the contributions meet all stakeholders' needs \cite{stol2014key}.

Our previous study\cite{chen2022reuse} investigated the implications of contemporary reuse in a medium-sized company. We identified that scaling InnerSource adoption could address the top identified challenges related to discoverability and ownership of the reusable assets, knowledge sharing, and reuse measurement. 
\paragraph{\textbf{Research gap}} 
In our previous study\cite{chen2022reuse}, we investigated the contemporary reuse process in a medium-sized company that was at the beginning of the software reuse journey. The company was in the initial stages of transitioning to microservices and implementation of InnerSource practices. Furthermore, the scale of reuse was not as large as the context investigated in this study. In this study, we investigate contemporary reuse in a company that is further along in the reuse journey and practices reuse on a large scale. Although previous studies found that microservices and InnerSource facilitate software reuse, the  benefits, costs, and challenges of reusing microservices in a cloud-native environment while adopting InnerSource practices are not yet investigated. 
In this work, we performed an in-depth case study using interviews, group discussions, company reuse-related portals, and company documents to investigate the implications of large-scale microservices reuse combined with InnerSource practices.

\section{Research Method - Case study design} 
\begin{figure*}
    \centering
    \includegraphics[width=\textwidth]{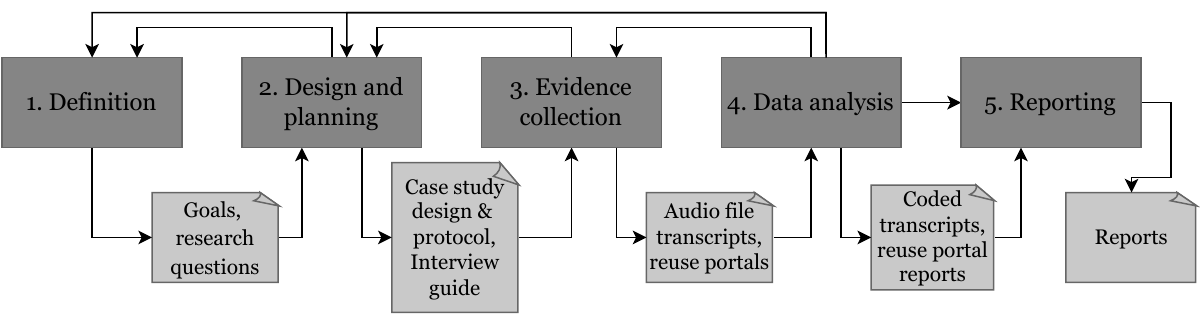}
    \caption{Research methodology followed in this study}
    \label{fig:RM}
\end{figure*}

\todo[inline]{DEB} \label{sec:rm}
Ericsson developed an ecosystem (details in Section \ref{sec:case}) for collaboratively developing common functions as microservices - which any application within Ericsson can reuse. We conducted a longitudinal case study to:
\begin{itemize}
    \item gain a deeper understanding of Ericsson's ecosystem,
    \item investigate the implications of contemporary reuse regarding its costs, benefits, challenges, and potential improvements, 
    \item revisit the challenges and improvements implemented by Ericsson to mitigate the challenges.
\end{itemize}
We followed the exploratory case study guidelines \cite{runeson2012case}, and for the data collection, we conducted semi-structured interviews, organized group discussions, and reviewed the company's reuse portal and documents. The case study guidelines consist of five interrelated phases. Figure \ref{fig:RM} shows the study phases and artifacts. 
We explain the steps followed in Phases 1-4 in Sections \ref{Sec:RMGoals} to \ref{Sec:RMAnalysis}. 

\subsection{Definition of goals and research questions} \label{Sec:RMGoals} 

In the first phase, we defined the scope of the study. This study is part of a research project with an overall goal of improving the internal reuse practices of the industry partners in the project. We decided on the scope of this study by reviewing the literature and considering input from the company representatives via a group discussion. All the authors and three practitioners from Ericsson participated in the group discussion to brainstorm the idea and scope of the study. In the brainstorming group discussion, we decided to investigate the current state of the reuse practice as a first step toward the project goal. We were interested in identifying the costs and benefits of practicing the internal reuse of microservices, as well as what challenges practitioners face and how to overcome these challenges. The following research questions are answered in this study:
\begin{enumerate}

    \item[RQ1:] What are the costs and benefits of practicing large-scale reuse of microservices? 
    \item[RQ2:] What are the challenges in practicing large-scale reuse of microservices, and how to overcome these challenges? 
\end{enumerate}
With RQ1, we aimed to identify the impact of the current reuse practice in terms of costs and benefits. In RQ2, we aimed to identify the challenges and suggestions for further improving the contemporary reuse practices at Ericsson. The intention is to investigate the challenges over time and understand the effect of implemented improvement suggestions in mitigating the challenges. 

We used multiple methods to collect evidence. Table \ref{tab:RQs_datacollection} describes methods we used for evidence collection to answer all research questions.

\begin{table*}[h]
\centering
\caption{Evidence collection methods used and their purpose}
\label{tab:RQs_datacollection}
\begin{tabular}{p{2.5cm}p{9cm}}
\toprule
\textbf{Method} &\textbf{Purpose}                             \\ \midrule

Group discussions & We conducted group discussions to define the scope of the study, validate interview results, and revisit the results. Details of the group discussion execution are provided in Section \ref{Sec:RMDesign}.\\

Interviews & We conducted 12 semi-structured interviews (details about interviews are described in Section \ref{Sec:RMDesign} and \ref{Sec:RMCollection}). \\

Reuse-related portals & The case company gave us access to their two reuse-related portals (details in Section \ref{Sec:RMDesign}). These portals provide information (e.g., rules and guidelines related to reusable assets i.e., common microservices) and data (e.g., related to the number, status, and maturity of the common microservices, other related metrics, and visualizations) that helped us to complement and triangulate the data collected through interviews.  \\

Company documents & We reviewed documents to corroborate the information provided in the interviews. The documents were identified and provided by the company representative. \\

\bottomrule
\end{tabular}
\end{table*}
\subsection{Design and planning} \label{Sec:RMDesign} We prepared the interview guide, identified the interviewee candidates, and identified the reuse-related portals in this phase.\\
\textbf{Group discussions:} We conducted multiple group discussions during the study. The first such discussion (involving three practitioners and all the authors) was used to brainstorm the scope of the study. Later, we conducted another group discussion (involving four practitioners and all the authors) wherein we presented and confirmed the preliminary results of our analysis. The second group discussion helped us clarify and understand the results related to all research questions. After a year, we conducted a third group discussion to revisit the challenges and improvements implemented to mitigate the challenges. The third group discussion involved all authors and two practitioners responsible for implementing the improvements. \\
\textbf{Interview guide:} The first two authors prepared the interview guide based on the research questions\todo[inline]{link to the interview guide?}. We conducted a pilot interview to validate the interview guide. There were no significant changes as the interview guide was on a generic level, given the semi-structured and open-ended nature of the interviews. \\
\textbf{Sampling:} Together with the company representatives, we identified the interview profiles in a brainstorming session. The interviewee profiles included all roles involved in software reuse as follows -
\begin{enumerate}
 \item[\textbf{P1}] Producers of common microservices. E.g., Developers and architects who contribute to common microservices. 
    \item[\textbf{P2}] Consumers of common microservices. E.g., Developers who reuse assets and potential (new) consumers of common microservices. 
    \item[\textbf{P3}] Software architects involved in reuse-related activities. 
    \item[\textbf{P4}] Relevant program and line managers.
    \item[\textbf{P5}] Other relevant roles related to promoting reuse practice at Ericsson.
    \item[\textbf{P6}] Mid-manager who owns the budget and approves the resources needed for different initiatives including developing for or with reuse.
\end{enumerate}
We prepared a one-page document\todo[inline]{link to the one pager} describing the purpose of the review study for practitioners participating in our study. One company representative identified one or more relevant candidates for each interviewee profile and sent an internal email to the interviewee candidates along with the one-pager. He also allowed the interviewee candidates to opt out before sharing their contacts with us. None of the interviewee candidates were subordinates to the company representative and could refuse to participate without any pressure. \\
\textbf{Reuse-related company portals}: The company representatives also provided access to two reuse-related portals: Continuous Integration and Deployment (CI/CD) and the marketplace portal. The CI/CD portal is a live portal that provides visual data on the different connection points between each microservice and the teams that use it. It also provides a live update on the pipeline execution status. The marketplace portal provides a list of all common microservices and includes several guides on creating common microservices. It provides detailed information on each microservice, such as the maturity level, reusability level, the team, including owners, and the list of contributors. The marketplace portal also displays a list of top contributors which is updated every 24 hours. The portals are discussed further in Section \ref{sec:case}.\\
\textbf{Company documents}: Some interviewees referred to the company's vision and processes during the discussions. To clarify the discussion points, we requested and received company documents and an audio-recorded presentation that clarified the company's strategy and processes related to developing, maintaining, and integrating the reusable assets. We used these documents to corroborate further the evidence collected through other methods. 
\subsection{Evidence collection} \label{Sec:RMCollection}
As mentioned in Table \ref{tab:RQs_datacollection}, we used multiple methods to collect evidence. After conducting the first group study to decide the scope of the study, we conducted semi-structured interviews. 

We conducted the interviews in the year 2021 through Microsoft teams.  13 candidates were identified and approached as relevant potential interviewees, out of which 11 responded. We also iterated the study protocol based on the evidence collected and analyzed. For example, we added the mid-manager: strategic project manager as an interviewee profile based on the information shared in some of the interviewees. In total, we conducted 12 interviews, as shown in Table \ref{tab:Interviewees}. \todo[inline]{More information is needed on how each role is related to reuse. Perhaps we can explain in Section 4?} The interviews lasted between 60 to 90 minutes. The three authors participated in all interviews alternating the roles of a lead interviewer, and notes taker and asking additional questions for clarification or completeness. All interviews were audio-recorded after obtaining consent from the interviewees. We divided the word-to-word transcription of the audio recordings among us. We also kept track of the timestamps of the audio file in the transcripts, which allowed us to listen to certain parts of the interview again if needed during analysis. We saved the audio recordings and the transcripts in a password-protected folder with access limited to the authors of this paper to ensure confidentiality. To anonymize the interviewees, we used acronyms instead of real names.

During the interviews, we realized that the state of reuse practices in the company is not in a fixed or finished state. Some initiatives were already either being planned or implemented to further improve the large-scale reuse of microservices. Thus, we decided to do a follow-up so that we can understand and present a relatively holistic view of the company's journey to improve and practice the reuse of microservices using DevOps and InnerSource practices. Thus, in the year 2022, we revisit the results through another group discussion. After taking consent from the participants, we recorded the discussion and transcribed the recording. 

The evidence collection from reuse-related portals was straightforward. We exported the data into excel files and stored it in a password-protected folder. In addition, we requested access to some documents that the interviewees mentioned. We reviewed the documents to gain additional information and support the claims made by the interviewees.

\begin{table}[]
\caption{Details of the interviewees}
\label{tab:Interviewees}
\begin{tabular}{@{}lcl@{}}
\toprule
\textbf{Profile} &\textbf{Experience*} & \textbf{Role}                                \\ \midrule
Producer(P1)       &  17yrs     & \ADPCA                            \\

Producer(P1)         &   13yrs    & \ADPPO                             \\
Consumer (P2)    & 12yrs      & \PlatA{}              \\
Consumer (P2)         &   4yrs    & \BAM                                \\
Architect and Consumer (P2,P3) &   17yrs    & \AppTL                                 \\ 
Architect and Consumer (P2,P3)&  10yrs     & \AppCM                  \\ 
Manager (P4)        &  17yrs     & \ADPPM                                     \\
Manager (P4)         &   16yrs    & \ADPHL                                      \\
Manager (P4)     & 20yrs      & \PlatM                             \\
Promoting reuse (P5)         &  15yrs     & \ADPA                                      \\

Promoting reuse (P5)         &  22yrs     & \ADPHead                             \\
Mid-manager (P6) &  14yrs     & \AppSPM                                            \\\\
\multicolumn{3}{l}{ADP refers to a centrally funded unit (more details in Section \ref{sec:case})}.\\
\multicolumn{3}{l}{\footnotesize * Experience at the time of conducting interviews is mentioned}.\\
\bottomrule
\end{tabular}
\end{table}
\subsection{Data analysis} \label{Sec:RMAnalysis}
\begin{figure}[!h]
    \centering
    \includegraphics[scale=0.7]{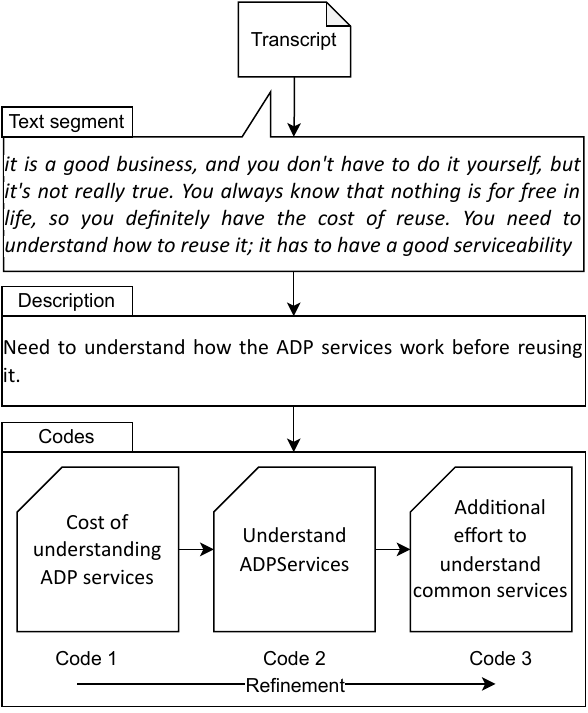}
    \caption{Example of the coding process}
    \label{fig:codingExample}
\end{figure}

\textbf{Coding:} The first two authors were involved in the analysis of the evidence collected through interviews, group discussions, and from reuse-related portals and documents. We coded the transcripts by extracting the relevant text segments to the research questions, describing the text segment in our own words, and labeling them with codes. To code transcripts consistently the first two authors conducted a pilot coding of one interview. After the independent coding, we discussed the individual codes where we revisited each text segment and coding and generated a code book in consensus. The code book contains the code name and its description. The first two authors continued to code the remaining transcripts using the coding guide. Through this process, we refined the code book in three iterations where the codes were either reformulated, combined with other codes, or separated as required. We also maintained a change log where we tracked the reasons for change. The code descriptions are updated whenever there is a change in the code. Figure \ref{fig:codingExample} provides an example of the coding process. 
\\\textbf{Abstracting and grouping:} In this process, the codes were categorized into higher themes. The first two authors grouped the codes that are semantically similar into themes. We performed the grouping in iterations where the themes were reformulated and merged with or separated into other themes. We presented the analysis results to the company representatives, and they acknowledged and confirmed our results. \\
\textbf{Analysing data from reuse portals:} Based on the information provided in the interview, we retrieved data from the reuse portals. For example, the interviewees mentioned the contribution to common assets. We exported all the contributions over the year 2021 into an excel sheet. We visualized the data into different charts and discussed the results with the company representatives to reflect on the reasons behind the patterns. 


\section{Case Description}\label{sec:case}
Ericsson is the case company, which is a large multinational company developing ICT (Information and Communications Technology) related products for service providers around the world.

Ericsson is transitioning towards a cloud-native approach to developing software applications and microservices. The company has developed an ecosystem to support such a transition to the cloud-native approach, wherein common functions are developed as microservices that can be widely reused by all applications within Ericsson. The ecosystem is referred to as the Application Development Platform - ADP ecosystem. The ADP ecosystem is responsible for providing the support - including technology, tools, practices, and principles - for the large-scale development and reuse of common microservices across Ericsson in a cloud-native context. The ADP ecosystem is our unit of analysis. The development of the ecosystem started in 2017 and it has continued to evolve since then. The ADP ecosystem consists of several aspects, which have been described recently in an article \cite{ADP}. In this study, our focus is to understand the effects of practicing large-scale reuse of common microservices in terms of observed costs and benefits, as well as challenges being faced and potential improvement suggestions to address them. To understand the results and discussions of the present case study, it is necessary to understand the context and background in which the reuse of microservices is being practiced at Ericsson. Therefore, we describe different aspects of the ADP ecosystem in the following sub-sections.

\subsection{Architectural framework}

\begin{figure}
    \centering
    \includegraphics[width=\textwidth]{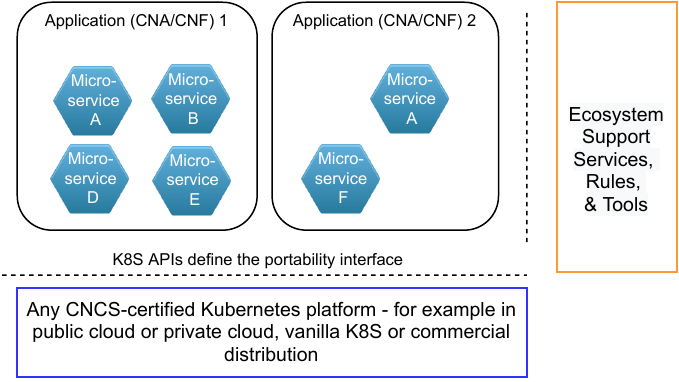}
    \caption{ADP Architectural framework}
    \label{fig:ADPArchitecturalFramework}
\end{figure}

The ADP ecosystem has adopted a modern cloud-native architectural style of microservices and containers. Kubernetes\footnote{https://kubernetes.io/} is used as the container orchestrator with  Helm\footnote{https://helm.sh/} as the package manager. Cloud Native Computing Foundation (CNCF)\footnote{https://www.cncf.io/} maintains a Kubernetes certification program to ensure conformance and interoperability across different vendor solutions - that have to support the required APIs. The microservices and applications are deployable on \textcolor{black}{any CNCF-certified K8S distribution}. The overall ADP architecture framework is depicted in Figure \ref{fig:ADPArchitecturalFramework}.

The common microservices correspond to different functional areas - such as security, data, network, monitoring, and management functions. These microservices are available for integration and reuse by all Cloud Native Applications (CNAs) across Ericsson. Most of these microservices are based on open source projects (e.g., various CNCF projects) - with their life cycle and configuration management customized to align them with the needs and requirements of the ecosystem. Ericsson teams not only adopt open source CNCF projects for internal use - but they also make upstream contributions with bug fixes and features to these open source projects including the Kubernetes project.

\begin{figure}
    \centering
    \includegraphics[width=\textwidth]{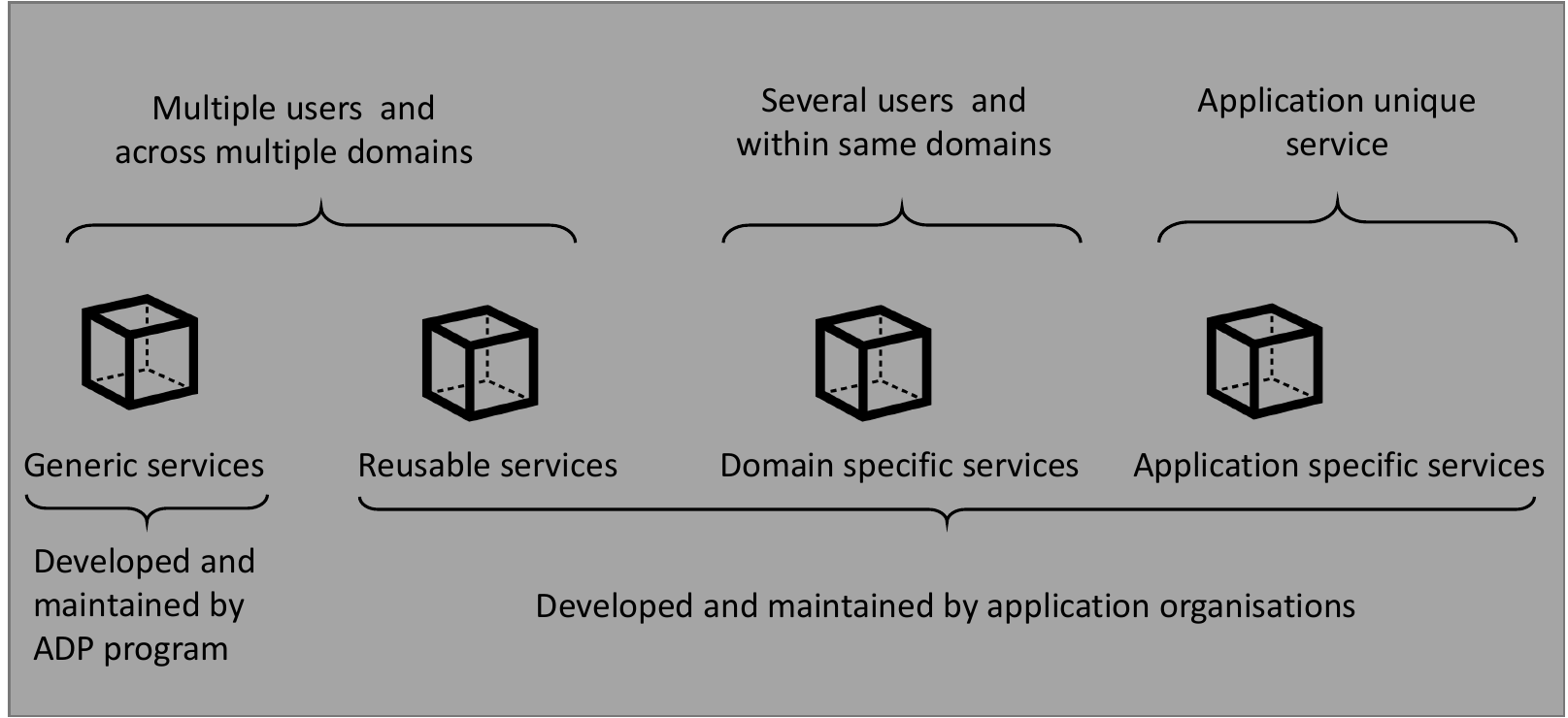}
    \caption{Different types of microservices in the ADP ecosystem}
    \label{fig:ServiceTypes}
\end{figure}

The microservices in the ADP ecosystem are classified into four types based on the two following two criteria:

\begin{itemize}
    \item Who are their potential users? Some microservices are more general than others - which means they have users from across different domains within Ericsson (see Generic and Reusable services in Figure \ref{fig:ServiceTypes}). We refer to such microservices as common microservices. On the other hand, some ADP microservices are only suitable for a specific domain, or even a specific application (see Domain and Application specific services in Figure \ref{fig:ServiceTypes}).
    \item Who is responsible for developing and maintaining them? Some common microservices are centrally funded (see Generic services in Figure \ref{fig:ServiceTypes}). The Generic services are owned by central teams that are part of the central organization - ADP program (see Section \ref{sub:ADP_program} for more details). The ADP microservices, other than the Generic services, are developed and maintained by the relevant application organizations.
\end{itemize}

The large-scale reuse of common microservices across a big organization like Ericsson implies the involvement of a large number of stakeholders - i.e., individuals and teams from different organizational units. These stakeholders participate in the ecosystem by developing, reusing, and contributing to the common microservices. The ecosystem has established some ground principles and rules to avoid inconsistent behaviors and implementations. These Design Rules (DRs) need to be followed by all microservices that wish to participate in the ecosystem. An example DR could include requirements for consistently naming and versioning the images or for consistently structuring the logs and metrics. Such an initiative to scale reuse at a large scale needs a high degree of automation. Therefore, the ecosystem also established a family of DR checkers that support developers and teams to automatically check the compliance of their microservices to these DRs. The DR checkers can be integrated into the CI pipeline by developers to test the compliance of their microservices.

\subsection{Marketplace}\label{Sec:Marketplace}

\begin{figure}
    \centering
    \includegraphics[width=\textwidth]{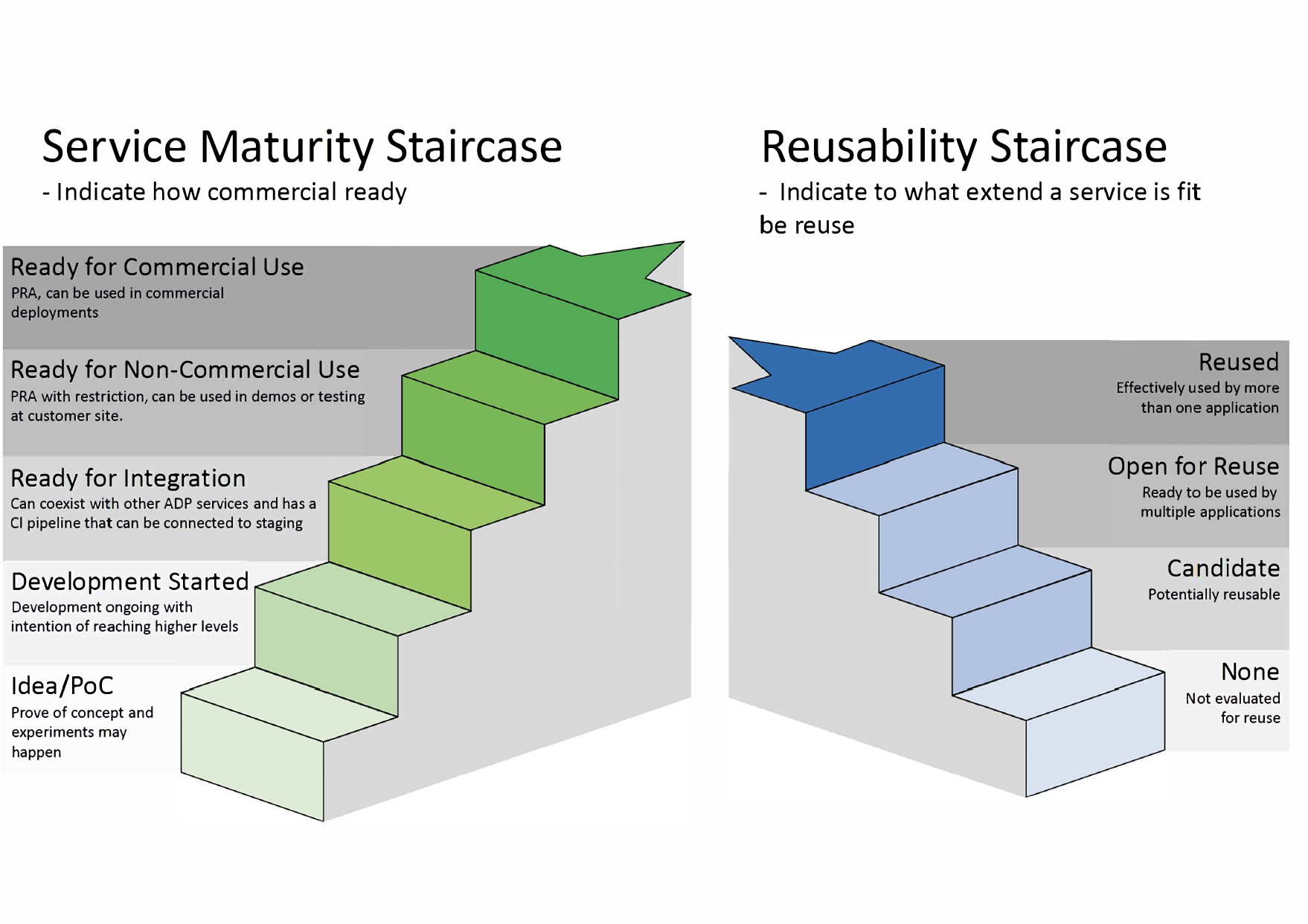}
    \caption{Maturity and reusability staircases (Reused with permission from Ericsson)}
    \label{fig:Staircases}
\end{figure}

The ADP ecosystem has developed an internal portal - known as ADP Marketplace - where all necessary information about available microservices has been made available. The Marketplace lists all microservices that are available for reuse and contributions. The Marketplace provides search and filtering options to quickly find the relevant microservices - e.g.; there are filters to group microservices on their type (i.e., Generic, Reusable, Domain specific and Application specific services). The ecosystem has also introduced two additional mechanisms for classifying the microservices (see Figure \ref{fig:Staircases}):

\begin{itemize}
    \item Maturity: The microservices maturity indicates how commercially ready a microservice is. At the very basic level, it is just an idea or a proof of a concept. The microservices at the highest level of maturity are the ones that are commercial ready. One of the main factors that separate mature microservices from relatively immature ones relates to compliance with the design rules. The ecosystem has assigned a different set of design rules to different maturity levels. Moving up the maturity level requires compliance with more design rules. Such an incremental way to attach an increasing number of design rules to higher maturity levels has enabled the ecosystem to clearly describe and communicate a milestone driven pathway for microservices to achieve maturity. 
    \item Reusability: The reusability staircase indicates how extensively microservices have been reused by different applications across Ericsson. At the basic level, there are those microservices that have not been reuse tested so far. At the highest level, there are those services that have been successfully reused by many applications.
\end{itemize}

On the one hand, the maturity and reusability staircases provide a mechanism for the microservice teams to see what is expected from them if they want to further improve their microservice. On the other hand, the staircases enable the microservice consumers to see what could be expected from different microservices.

The Marketplace also provides guides and tutorials about using, developing, and contributing to the microservices. The Marketplace hosts over 280 microservices, of which over 50 have the highest service maturity (see maturity levels in Figure \ref{fig:Staircases}). For each microservice, the Marketplace includes the necessary information (e.g., the owner team, the list of contributors from outside the owner team) and links (e.g., to the relevant code repository) that are important for the installation and contributions to the microservice. 

\subsection{DevOps - Continuous Integration and Delivery (CI/CD)}

The ADP ecosystem has developed a CI/CD strategy to ensure the continuous development and testing of microservices and their subsequent reuse by consumer applications. The CI/CD strategy involves two connected but independent pipelines: microservices' own CI pipelines and applications' (that consume the microservices) assembly pipelines. The microservices have their CI pipelines wherein the microservices teams continue to release new versions. The applications that consume the microservices have their own pipelines for continuous assembly and release of applications. The microservices CI pipelines are connected with the relevant applications' assembly pipelines using Spinnaker pipelines. When a new version of a microservice is released by the service team, it goes to the staging environment of the relevant consumer applications for them to test if they want to replace the older version. This way of decoupling the microservices CI pipelines from the applications' assembly pipelines is an important enabler for the large-scale development, testing, and reuse of microservices across Ericsson.

\todo[inline]{We need to include the updated numbers at the end - the updated numbers shall include the following: \\

The number of available microservices in the marketplace\\
The number of connection points\\
The number of microservices with the highest reusability and maturity etc.}

\subsection{InnerSource practices}

The common microservices have been used in several applications - that frequently ask for new requirements and enhancements. Such an arrangement wherein a large number of consumer applications continue to ask for more features have the potential to become a bottleneck. The central teams would not be able to implement all changes in time - resulting in delays and bottlenecks. To overcome these delays, the ecosystem introduced InnerSource as a collaborative way of maintaining the common services wherein the consumers of the services are not using the services, but they also contribute with bug fixes and enhancements. Taking inspiration from the InnerSource Commons \cite{ISCommons_IP}, the ecosystem has introduced some InnerSource practices, roles, and responsibilities to govern the development and maintenance of the microservices. 

The common microservices are handled as InnerSource projects wherein each service has a guardian, one or more trusted committers, and a product owner. A \textbf{guardian} is the technical owner of the service, who is responsible for developing the contribution guidelines, reviewing the contributions, creates and participating in the discussion forums for the service. A \textbf{trusted committer} is someone who has been contributing to the service for a while and therefore has developed a good technical understanding of the service. There can be multiple trusted committers for a service. The trusted committers of service play an important role in developing a sustainable community for an InnerSource project. They can mentor the new contributors and participate in the discussion forums. The trusted committers are usually from outside the unit that developed the service, while the guardians are usually from the team that originally developed the service. Lastly, each service also has a \textbf{product owner}, who is responsible for managing the backlog and roadmap for the service. The source code and backlog of the common microservices are open to all within Ericsson to encourage InnerSource contributions. However, the contributions are only merged after they are reviewed by the relevant guardians or trusted committers.

\subsection {ADP Program}
\label{sub:ADP_program}
The ADP Program is the central organization that is responsible for sustaining and governing the ADP ecosystem. The ADP Program includes the central teams that are responsible for the development and maintenance of over 50 centrally funded Generic services that are reused extensively in different applications across Ericsson. The technical leadership of the ecosystem is provided by an architecture team. The architecture team is part of the ADP Program. It drives the design and maintenance of the principles, DRs, and guidelines on the cloud-native approach to developing microservices and applications. There is also an architectural council wherein the ADP architects meet with the architects from the applications that consume common microservices to discuss the future directions and changes to the guidelines and DRs.

The ADP Program also includes a unit - referred to as ADP Anchor - which is responsible for designing policies and guidelines for the ecosystem. The ADP Anchor unit continuously updates the support for developing, reusing, and contributing to the common microservices. The ADP Anchor unit includes over 100 practitioners who are performing different anchoring functions within the ADP ecosystem. The ADP Anchor unit also keeps track of the progress on different initiatives by collecting and analyzing different metrics such as the number of contributions, top contributors, the maturity level of different microservices in terms of DR fulfillment, and which microservices are more/less successfully integrated by consumer applications, etc.

\section{Results RQ1: Software reuse costs and benefits}
\label{sec:resultsRQ1} 
\todo[inline]{DEB - need to finalize roles and update them in the results}





In this section, we discuss the costs and benefits of the large-scale reuse of microservices using DevOps and InnerSource practices in a cloud-native environment at Ericsson. We refer to the microservices that are used by multiple users and across multiple domains (see Generic and Reusable services in Figure \ref{fig:ServiceTypes}) as common microservices.

\begin{figure}
    \centering
    \includegraphics[width=\textwidth/2]
    {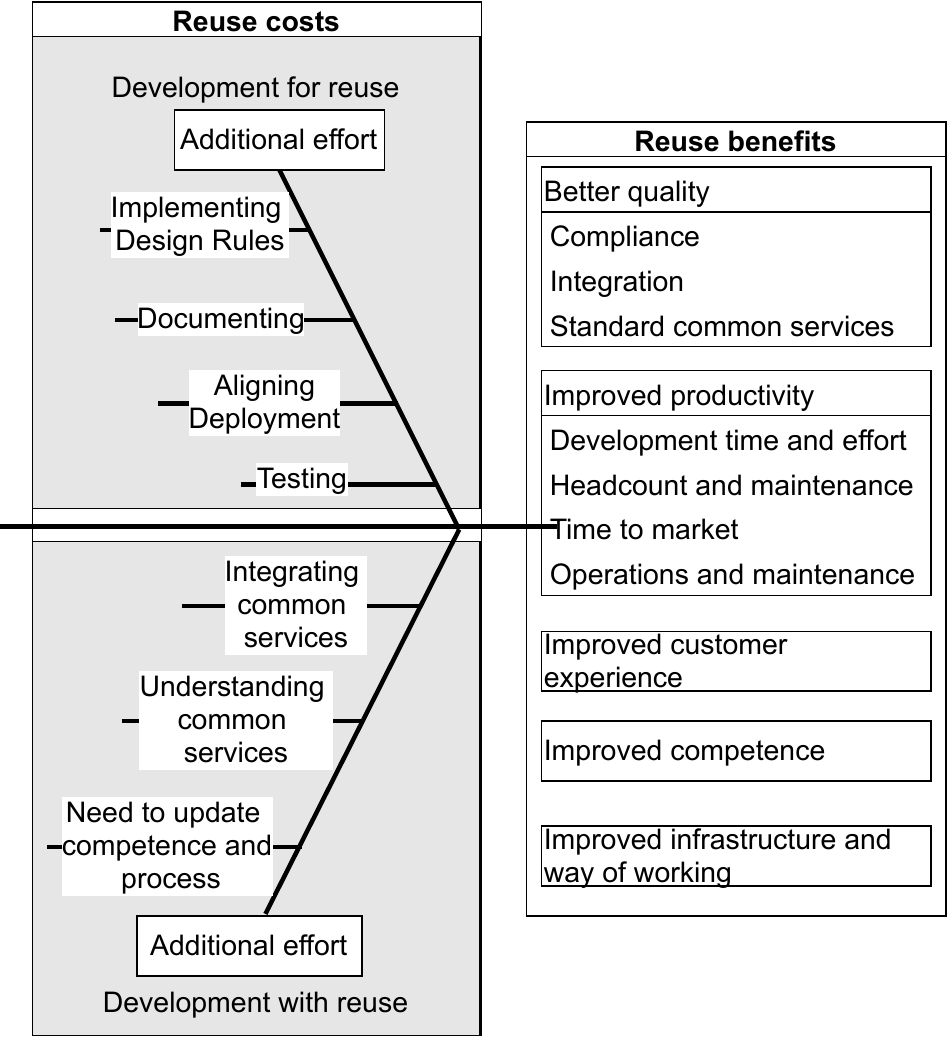}
    \caption{Costs and benefits involved in contemporary reuse.}
    \label{fig:costs}
\end{figure}

The main purpose of software reuse is to reduce costs and increase quality. However, some upfront costs will inevitably occur when developing common microservices, which will pay off when consumers reuse such microservices in their applications. On the other hand, using common microservices also comes with some additional costs, which an interviewee summed up: \textit{\say{there is no free lunch}}. In our study, we refer to costs as the additional effort required to develop and integrate common microservices. We classified the identified costs into \textit{\say{development for reuse}} and \textit{\say{development with reuse}} and grouped the benefits into themes when applicable, as depicted in Figure \ref{fig:costs}. We describe the different software reuse-related costs and benefits in this section.
\subsection{Software reuse costs} 
\label{Sec:Costs}
In this sub-section, we discuss the identified software reuse costs from two perspectives - costs involved in development for reuse and development with reuse.\\
\subsubsection{Development for reuse} In this sub-section, we discuss the additional effort practitioners perceive they spent in the development for reuse.\\
\textit{\underline{Additional effort in implementing design rules:}} Ericsson has developed design rules (DR) that describe requirements for handling issues such as configuration, integration, and ensuring backward compatibility. \textcolor{black}{All microservices developed in Ericsson need to conform to the DRs. However, there is an increased need to ensure conformance to the DRs in case of common microservices which are integrated into different applications across Ericsson.} As mentioned in Section \ref{sec:case}, it is important to ensure that the common microservices are compliant and developed consistently. Implementing DRs and compliance requirements such as security takes considerable time, according to an \ADPPO of a common microservice. In addition, a Tech lead involved in producing and consuming a common microservice referred to implementing DRs as \say{\emph{a painful and limiting process}}. The Tech lead continued the discussion by highlighting that such additional costs are necessary to make the services reusable.  \textcolor{black}{The DRs are also essential to get the alignment that allows for a common look and feel for the customer}. Thus, DRs are an investment necessary for getting the benefits in the future (discussed in Section \ref{sec:benefits}). \textcolor{black}{Furthermore, it is important to note that the additional effort in implementing design rules is not specific to common microservices. However, there is an increased need for \emph{policing} conformance for common microservices compared to non-reused microservices where the developers can decide on the \emph{pace} of conforming to the design rules. } \\
\textit{\underline{Additional effort in documentation, deployment and testing:}} In addition to the code itself, other artifacts, such as its associated documents and test cases, require additional effort to make them reusable. One of the developers mentioned that common microservices required additional documentation because the code needs to be understood by its users. Different teams use common microservices with varying deployment environments. Therefore according to a \PlatA the common microservices \say{\emph{need to be adapted to support the deployment}}.
\subsubsection{Development with reuse} In this sub-section, we discuss the additional effort practitioners perceive in the development with reuse.\\ 
\textit{\underline{Understanding and integrating common microservices:}} A \AppSPM mentioned that before reusing the common microservices, the teams need to understand the service and how to integrate it. In addition, the SPM mentioned that the common microservices are developed for a cloud environment. Not all applications are fully cloud-native yet. Even to use a small microservice, the team needs to update their infrastructure to a cloud environment and set up containers and CI/CD pipelines. According to a \AppSPM\say{\emph{It adds a big footprint to customer deployment}}. However, it is a one-time setup, and once established, the integration becomes easier, as stated in Section \ref{sec:benefits}. Integrating each common microservice also adds a few dependencies and requires additional integration testing. A \PlatM mentioned that \say{\emph{reusing stuff adds dependencies...and that can slow you down... that is also why people tend to do it themselves.}} \textcolor{black}{The additional dependencies are an inevitable consequence of a microservice-based architecture where a single microservice should have a single responsibility thereby resulting in several microservices and the dependencies between them.}\\
\textit{\underline{Need to update competence and process:}} Common microservices are often up to date with the changing technology, which implies that the users need to constantly update their competence to continue to use the common microservices. Similarly, although there is an upfront cost, the knowledge and competence gained can be used \say{\emph{in their[users'] own development and any kind of reuse competence}} according to a Tech lead. 

As seen, most of the costs when developing and reusing the common microservices lead to benefits in the long run. We discuss the identified benefits of reusing common microservices in Section \ref{sec:benefits}.

\subsection{Software reuse benefits} \label{sec:benefits}
In this sub-section, we discuss the software reuse benefits identified in our study.
\subsubsection{Improved quality}
Most interviewees shared the view that the quality of the common microservices is generally better. In addition to the ADP program team, the applications and the platform teams also highlighted this benefit. A \AppTL mentioned that: \say{\emph{the number of issues that I have to discover will be lower because all the other applications already used this product [common microservice]. It reached a certain amount of maturity and quality before getting there.}}
The \ADPPM reflected that the improved quality results from a good quality assurance process considering the feedback from microservice users. The \ADPPM added that: \say{\emph{[common microservices] has more checks, has a bit more pressure from the design rules and the test coverage perspective.}}
Below we describe the specific attributes of quality that the interviewees discussed in the interviews -\\
\textit{\underline{Better compliance}}:  When the microservices become available for reuse in the organization, they are already compliant with some, if not all, of the company's compliance requirements. One of the \PlatA stated that most of the common microservices are adopted from open source, and they are complaint to the company's directives: \say{\emph{We made them [common microservices] more secure...more suitable for telecom's environment and..more compliant to Ericsson's directives ... like the Ericsson's security requirements.}}\\
\textit{\underline{Better integration with products}}: Ericsson has multiple product offerings, and common microservices often need to integrate with other products. The products may use different technology or programming language, causing integration issues. The  \AppCM confirmed that common microservices support such integration - \say{\emph{If you are reusing some reusable component or standard interface, it is much easier to integrate.}}
    A \AppTL  mentioned the reason for easy integration as \say{\emph{I think it’s a lot easier for other products to quickly pick up [mentions a specific microservice] and get a deployment of that in their product.. based on both guidelines from ADP ecosystem and the experience from us using Cassandra}}. \\
\textit{\underline{Standardized common services}}: The \ADPPO highlighted that using the same service for all the common functions in different applications provides uniformity. The \AppTL added: \say{\emph{That [reusbale microservice] is something we now can productify in a generic solution}}
\subsubsection{Improved productivity}\label{Sec:Productivity}
The interviewees shared that reusing services also improve productivity. Both the members of the ADP program and the application team perceive that productivity is improved as follows: \\
\textit{\underline{Saves development time and effort}}: The common microservices are developed once and reused multiple times. A \ADPPM highlighted that \say{\emph{it will be more efficient if we developed that [common microservices] only once and that can be reused by all the different applications [teams].}}\\
\textit{\underline{Reduced headcount and easy maintenance}}: Common microservices reduce headcount and provide easy maintenance. An \ADPPO responsible for one of the common microservices used across the organization mentioned that \say{\emph{There are only eight developers who are maintaining [mentions a specific microservice]. So the cost is consolidated to a small team. When any vulnerability is reported, this team will address it instead of going to many different people and maintaining different versions of logging.}}\\
\textit{\underline{Reduced time to market}}: The benefit of reduced time and effort results in faster time to market. An \ADPCA shared that common microservices are: \say{\emph{able to bring faster solutions to customers as well.}}\\
\textit{\underline{Facilitates operations and maintenance (O\&M)}}: Additional benefit from the platform team perspective is that the common microservices are used for conducting operations and maintenance activities across different applications. A \PlatM reflected \say{\emph{The fact that we could take in common microservices as a base instead of building something on our own. It is a massive benefit.}}

\subsubsection{Improved customer experience}
Both producers and consumers of common microservices mentioned the benefit from the customers' perspective. Common microservices improve alignment and provide customers with the same look and feel. According to a \PlatM reuse also benefits the customer installation process. If multiple applications use a shared service for certain functionality, then customers need to install it once and save time and effort for all products. Finally, reusing common microservices enhances customer learning. One of the \PlatA mentioned that \say{\emph{If they[customers] have multiple Ericsson products. If you have the same service in three different Ericsson products,... then once they[customers] know and learn how that service works, they understand all of the products.}}
\subsubsection{Improved Competence} By reusing services, the competence is consolidated to one team instead of distributed. One of the \ADPPO mentioned that \say{\emph{the competence is available in one team instead of having competence in each team}}. Internal specialized teams responsible for common microservices help in the ownership and management of shared assets. A \PlatM confirms that \say{\emph{Instead of trying to get implementation by a vendor or open source, it is a benefit that we have a team that we can work with it.}} In addition, according to an \ADPA when teams collaborate and contribute to other teams, \say{\emph{it improves the competence within the team and the organization as well.}}
\subsubsection{Improved infrastructure and way of working}
When reusing common microservices, the reuse occurs beyond software; it is also about reusing the ways of working and test framework, for example. A \ADPPM stated that \say{\emph{So it’s not only about [code] reuse, but also it is the reuse of the practice and ways of working, test framework, and many things}}.
\subsection{Contemporary reuse Cost-benefit tradeoff}
\textcolor{black}{As discussed in Section \ref{Sec:Costs}, some additional costs are involved in developing and using the common microservices. However, the study participants highlighted that the additional costs led to several benefits when the common microservices are reused, such as reduced development time, effort, and headcount (see Section \ref{Sec:Productivity}). Therefore, the additional costs for developing the common microservices are an investment whose returns are achieved as and when different users in the organization reuse the common microservices. 
The users of the common microservices avoid the additional costs as they do not have to develop from scratch. Cost avoidance is already achieved when a few users use the microservices. However, the cost avoidance is substantial when a significant number of users (e.g., 10+) use the common microservice. In addition, the users benefit from the advantages common microservices offer, such as providing customers with a common look and feel. Therefore, the benefits of reusing common microservices outweigh the costs incurred in developing and reusing them. }

\section{Results RQ2: Contemporary reuse challenges and improvement suggestions}
\label{sec:challenges}

This section discusses the practitioner perceived challenges and improvement suggestions in reusing common microservices that are used by multiple users and across multiple domains (see Generic and Reusable services in Figure \ref{fig:ServiceTypes})) within Ericsson. \textcolor{black}{In addition, we provide details on the improvements already implemented by Ericsson to address the challenges.} 

\subsection{Challenge 1: Bottleneck at the centrally funded teams} \label{sec:bottleneckchallenges}
\begin{figure*}[!t]
    \centering
    \includegraphics[width=\textwidth]{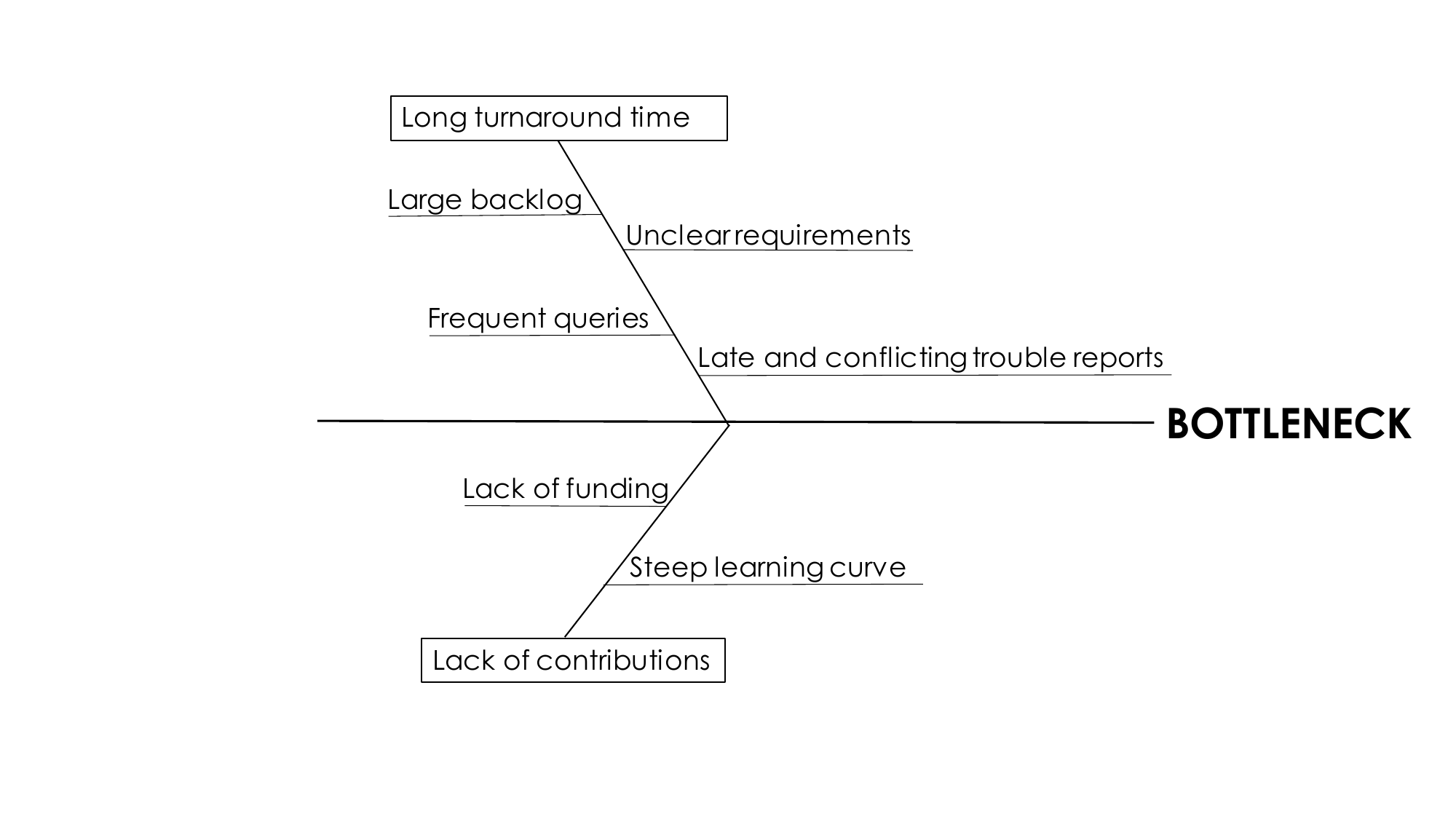}
    \caption{Causes resulting in a bottleneck.}
    \label{fig:bottleneck}
\end{figure*}
At the time of conducting interviews, there was a lot of dependency on the centrally funded teams to support the development and maintenance of the common microservices, which created a bottleneck. The reasons for the pressure on the centrally funded teams resulting in a bottleneck are depicted in Figure \ref{fig:bottleneck} and described in this sub-section.
\subsubsection{Long turnaround time} \label{Sec:turnaroundtime} The causes for long turnaround time are discussed in this section and depicted in Figure \ref{fig:bottleneck}. \\
\textit{\underline{Large backlogs:}} Due to the dependency on the centrally funded teams, they had large backlogs as they received requirements from many different teams. The new requirements need to align with the purpose and scope of the common microservices, and there is a possibility that the requests may not be accepted. According to an \ADPCA, \say{\emph{just providing indications if they would address the request, that can take probably, in some cases, more than a month or even more than that.}} In some cases when the waiting time is too long, teams may implement a workaround by themselves. The teams may decide to replace the workaround when the producers eventually implement the fix or the new requirement. However, some applications may continue to work with the workaround solution. Therefore, long waiting times can hinder teams from reusing the common microservices.\\ 
\textit{\underline{Lack of details in requirements:}} Lack of clarity and details in the requirements coming from the users of the common microservices also came up as a reason for a long turnaround time. A \ADPPM mentioned that  \say{\emph{sometimes the requirement is not well written, so we don’t really know what they want.}} \\
\textit{\underline{Frequent queries:}} The centrally funded teams received many queries on how to use the common microservices and how to migrate to cloud native solutions. According to a developer, the challenge is that \say{\emph{we do not have that much bandwidth to answer all the queries or help them end to end.}} \\
\textit{\underline{Late and conflicting trouble reports (TRs):}} According to a \ADPPM sometimes the TRs from consumers come at the last minute, which puts pressure. In addition, the \ADPPM added that \say{\emph{the main complaint that we [centrally funded teams] get sometimes is that this alignment with what they [consumer] consider is a TR, but what we think is new functionality.}} The misalignment could also be due to the misinterpretation of the design rules and generic product requirements and their applicability (discussed more in Section \ref{MisGPRDR}). 

\textcolor{black}{As the number of common microservices grows, the number of feature requests and queries will inevitably increase. The improvement suggestions provided by the interviewees to manage the increasing backlog were to encourage consumers to make contributions (discussed in Section \ref{sec:Innersource}). The implemented improvements to improve the turnaround time are described in Section \ref{Sec:ImprovementLongTurnaround}.}

\subsubsection{Lack of contributions} 
\begin{figure}
    \centering
    \includegraphics[width=\textwidth/2]{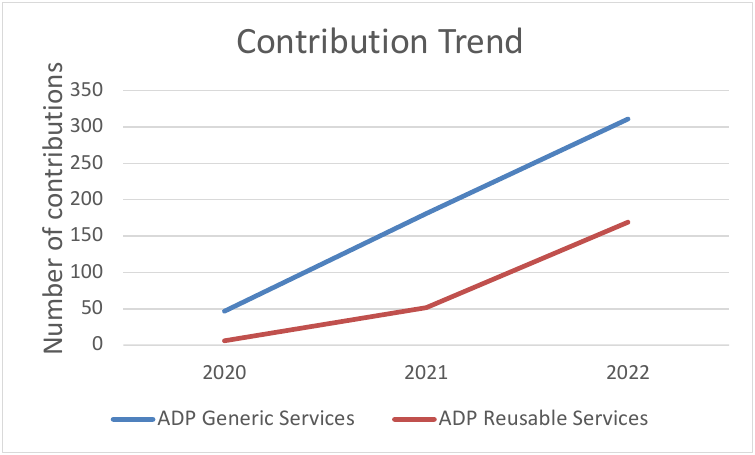}
    \caption{Contribution trend between 2020-2022.}
    \label{fig:ContTrend}
\end{figure}

Ericsson decided to promote InnerSource - a collaborative way of working wherein consumers consume the microservices and contribute to their development. However, an \ADPCA reported that \say{\emph{there have been too few contributions}}. The lack of contributions is evident from the data we collected from the marketplace portal. Figure \ref{fig:ContTrend} visualizes the contributions between the years 2020 and 2022. As seen in Figure \ref{fig:ContTrend}, the contributions were fewer in the year 2020 compared to 2022.  
\textcolor{black}{The applications have their own release plans and may question the potential benefits they will get by contributing to the common microservices.}
For example, the \ADPA mentioned that the consumer might think \say{\emph{what benefit I will get to build something that others will reuse... because I am spending money on this and why should I make it available for others}.} The managers promoting InnerSource want the applications to \textcolor{black}{adopt a more collaborative way of working}. For example, \say{\emph{if I am a consumer, I am taking it[reusable common microservice] and I am using it freely, then I should also contribute; otherwise, the ecosystem will not work.}} \\
\textit{\underline{Lack of funding:}} The \ADPHead mentioned that \say{\emph{The hard sell has been the thing about making my work reusable for others without getting any kind of payment for it.}} One of the \AppSPM mentioned that they need to prioritize customer requirements that customers are paying for. Contributing to common microservices may not result in immediate benefits in terms of revenue; therefore, they may not be prioritized over the requirements yielding immediate revenue benefits. \\
\textit{\underline{Steep learning curve:}} The applications may use different programming languages, tools, and test frameworks and need to understand the common microservice before they can start contributing. The overall \ADPPM mentioned the learning curve as a barrier to contributions - \say{\emph{not all the services are written in the same language, maybe they are different from the one that the application is using, but that could be a barrier for contributing.}}

\textcolor{black}{The InnerSource initiative at Ericsson has been a journey wherein the organization is adopting its policies and practices to address the identified challenges and further improve its ways of working. The practitioners suggested several improvements to address the challenges discussed above, which are detailed below in Section \ref{sec:Innersource}. In addition, Ericsson has also implemented some steps to increase the InnerSource contributions, which are elaborated in Section \ref{Sec:ImprovementContri}.}


\begin{figure*}[h]
    \centering
    \includegraphics[width=\textwidth]{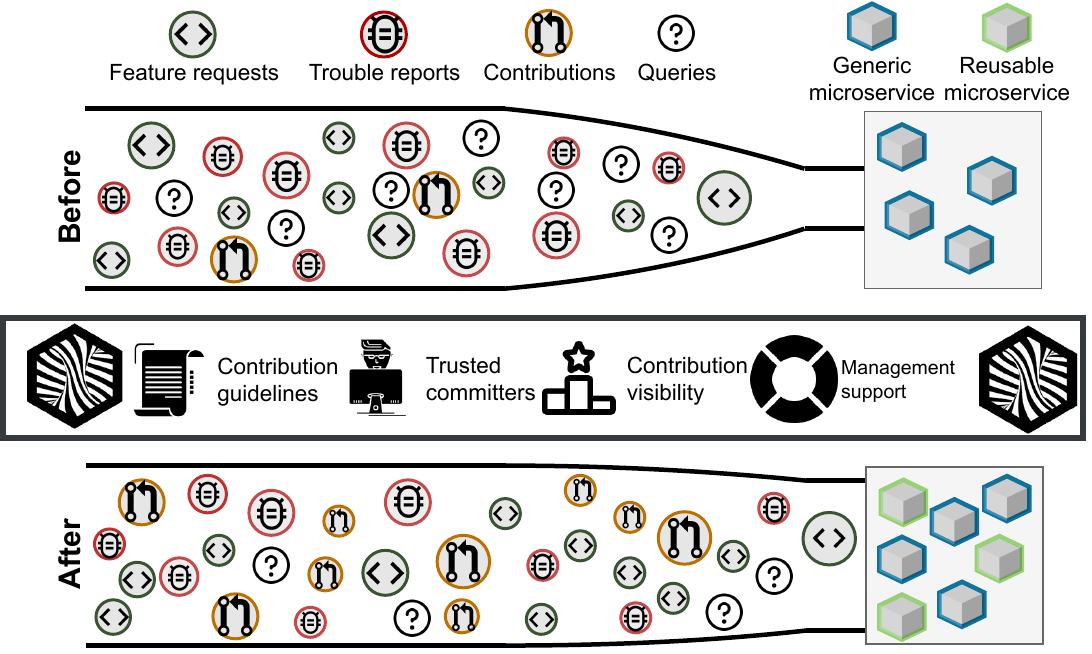}
    \caption{InnerSource as a reuse facilitator.}
    \label{fig:ChallengesToImprovements}
\end{figure*}
\subsection{Improvement suggestion 1: Scale InnerSource adoption} \label{sec:Innersource}To prevent the centrally funded team from becoming a bottleneck, Ericsson promoted the use of the InnerSource way of working. However, as mentioned earlier, only a few contributions were made at the beginning of this journey. The interviewees identified the need to scale the InnerSource adoption. The scenario before the improvements and the desired scenario after scaling InnerSource adoption are depicted in Figure \ref{fig:ChallengesToImprovements}. As seen in Figure \ref{fig:ChallengesToImprovements}, the number of contributions were fewer before scaling InnerSource adoption. In addition, only the centrally funded teams were responsible for providing the common microservices. Particularly the aim was to increase the number of contributions by implementing the following improvements: 
\subsubsection{Guidelines for InnerSource contributions} Both the producers and consumers perceive it important to improve the guidelines on how to practice InnerSource. The process for contributing should be defined as noted by \AppCM \say{\emph{If I want to contribute [to a common microservice], what do I need to do? Who do I need to contact? How will it get reviewed? Do I need to present the design document?}}. In addition, the microservice owner should describe guidelines for the service; for example, the \ADPHead stated that the owner \say{\emph{should describe the rules for the microservice. This is how I expect the code to be.... This is how you extend the test cases and the documentation.}}

The \ADPPM mentioned that there is a \say{\emph{need to make sure that [the contribution] is properly written, that is properly documented, that you also get the test cases to secure that part of the code is not broken later on.}} The interviewees also recognize the need to define a process for reviewing contributions. 
\subsubsection{Expand the IS contribution possibilities} As mentioned in Section \ref{sec:bottleneckchallenges}, one of the reasons for the bottleneck is the late and conflicting trouble reports. One way to expand the type of InnerSource contributions is to encourage the users to contribute to the resolution of the bugs - rather than only submitting the trouble reports and waiting for their resolution. In this way, the users can make smaller and more frequent contributions. The users have first-hand information about the issue; therefore, the problem resolution will take much less time than assigning the trouble report (TR) to developers from the team owning the microservice and waiting for the resolution. If the consumers do not have the resources to resolve the TR completely, they can still support producers in resolving the TR. For example, as stated by the \PlatM \say{\emph{Maybe then it's much easier to go to the owner and say I found this problem, this exact line of code, and this change will fix it.}} In summary, accepting smaller yet essential contributions in support of troubleshooting or resolution of TRs can encourage receiving more contributions and reduce the workload on the centrally funded teams in terms of feature requests and bug/issue reporting.

In addition to the different ways of contributing \textit{to} a microservice, Ericsson also encouraged contributing \textit{with} a microservice. The different types of contributions are described below - 
\begin{itemize}
    \item Contributing code - These contributions are changes in a microservice as a result of a feature request or bug fix. 
    \item Contributing microservices - A team besides the centrally funded team can make their microservice open for reuse. The team that contributes with a microservice is also responsible for the maintenance of the microservice. 
\end{itemize}
\subsubsection{Synchronise and plan contributions} The contributions need to align with the purpose and architecture of the common microservice. The contributors should first discuss the plan for contribution with the corresponding guardian. For example, the \ADPPM mentioned that \say{\emph{the new functionality that you[contributor] want to introduce in the product is aligned with the purpose of the product, is aligned with the architecture, so it’s ok for you to contribute this.}} The owner team is more prepared to review the contribution and respond quickly if the contributors have discussed their plan instead of directly making a large contribution without any heads-up.
\subsubsection{Innersource contribution visibility} As the goal is to encourage more contributions, one way to encourage contributions is to make good examples of contributions visible. The ADP marketplace already displays the top contributors. However, the \ADPHead recognized the need to make different levels of contributions visible on the ADP dashboard. He added \say{\emph{where somebody contributed a commit or few lines of code, I want to get those on a dashboard as well.}} 
\subsubsection{Management and financial support} The interviewees want to improve the mid-managers' (SPM) encouragement in making contributions to other projects. From the SPM perspective, an additional budget for contributing with a microservice could make it easier, \say{\emph{if we want them [microservices] to be better at security and footprint...you have to give the money to do it.}} In addition, the \ADPHead mentioned that \say{\emph{we need to get all of our top-level leaders on board with InnerSource.}} 
\subsubsection{Improve openness to contribute and receive contributions} Some interviewees mentioned the need to improve the behavioral aspects, such as openness towards providing and receiving contributions. The \ADPCA raised the importance of the teams being self-empowered. He added that the teams should take responsibility for contributing. From the contributors' perspective, \AppCM mentioned that the owners of the common microservices \say{\emph{need to be open to allow us to contribute.}} She added that the contribution process should be like open source projects where everyone can contribute.

\subsection{Improvements implemented to address the bottleneck at the centrally funded teams} As described in Section \ref{sec:bottleneckchallenges}, the reasons for the bottleneck at the centrally funded teams are due to the long turnaround time in addressing the requirements and lack of contributions. In this section, we describe The approaches Ericsson adopted to mitigate the bottleneck challenge.  
\subsubsection{Mitigating long turnaround time}\label{Sec:ImprovementLongTurnaround}
In our follow-up study, we found that the backlogs are still large; however, it is not a major challenge anymore due to the following two reasons:  \\
\textit{\underline{Not all requirements are part of the active backlog:}} The centrally funded teams mark the requirements they will not work on in the next 12 months as candidates for the future. Such requirements are not part of the active backlog. In this way, the consumers have an estimate of when their requirements will be considered and can decide if they would like to wait or make a contribution. This approach has also led to receiving important requirements from consumers, allowing the centrally funded teams to focus on prioritized requirements.  \\
\textit{\underline{Distributed responsibility:}} The consumer teams are welcome to contribute if the centrally funded teams cannot address the requirements on priority, therefore not adding pressure to the centrally funded teams and resulting in fewer management escalations. In addition, the consumer teams can have more control over the timeline by contributing. 
\subsubsection{Increasing contributions}\label{Sec:ImprovementContri}
As mentioned in Section \ref{sec:Innersource}, Ericsson encourages contributions \textit{to} a microservice (owned by the centrally funded teams) and \textit{with} a microservice (owned by application organizations). 

The lack of consumer contributions \textit{to} the microservice  puts pressure on the centrally funded teams, resulting in a bottleneck.
The lack of contributions to the microservice, as perceived by the interviewees, was mainly due to the lack of funding. The contributors sought management and funding support to be able to make contributions to the microservice. The top management has now set targets on the number of contributions expected from the teams. The teams are expected to contribute from their existing resources; this expectation is aligned with the InnerSource principles. 

Teams contributing \textit{with} a microservice may also have the same bottleneck challenge as the centrally funded teams. For example, once the team contributes \textit{with} a microservice and many other teams start reusing it, they may have to address many bugs and user feature requests from the teams using the microservice. The ADP anchor refers to this as \say{\emph{the Punishment Clause:}} getting punished with a lot of additional work for contributing with a microservice. The ADP anchor has initiated a funding program where the contributors get support to maintain their contributed microservice. If a team contributes with a microservice that is integrated into three or more products, then they will unlock benefits in terms of additional bandwidth to maintain their code (handle feature requests, queries, and bug fixes).

The interviewees identified the steep learning curve as another barrier to contributions. As mentioned in Section \ref{sec:bottleneckchallenges}, the common microservices are not implemented using the same technology. While the contribution guides have improved since 2021, the learning curve remains challenging.  Adding test cases is difficult when common microservices use different test frameworks. One workaround is to enforce using the same test framework for all the microservices. 

Overall due to the several initiatives taken by Ericsson, the number of different types of contributions: contributing (i) to microservice (ADP Generic Services)and (ii) with microservices (ADP Reusable Services) has increased over the years (see Figure \ref{fig:ContTrend}). 

\subsection{Challenge 2: Interpretation of design rules and generic product requirements} \label{MisGPRDR}
Generic product requirements (GPR) are compliance requirements such as security. All Ericsson products and applications need to comply with the GPRs. The interviewees reported that \textcolor{black}{when} the GPRs were \textcolor{black}{first specified, they were} specified on a high level, for example, for the product level and not on the microservice level. The applicability of the GPRs on the microservice level was unclear and created misunderstandings between the producers and consumers. 

Similarly, developers may misinterpret the design rules if they are not specific and detailed (see Section \ref{sec:case} for details on design rules). The lack of details made it difficult to check whether or not the design rules were implemented. In addition, different teams could interpret the design rules or generic product requirements differently if they are not specific enough. The \ADPHL mentioned that \say{\emph{So ADP [program team] has interpreted this generic product requirement or the design rules in one way and implementing like that. And our security people have another interpretation; I think this is a gap.}} From a consumer perspective, this challenge is of high priority as it resulted in delays in releases. A \AppSPM mentioned that \say{\emph{it became very obvious that we could not release the software due to several security gaps in the common microservices.}} It takes a long time to get the security fixes due to the centrally funded teams' long turnaround time, which further delays the release.

\subsection{Improvement suggestion 2: Improve understanding of design rules and generic product requirement}\label{IA:MisGPRDR} When implementing the design rules, it is important that the developers understand the importance of the design rules. An \ADPCA shared that developers need to know \say{\emph{what’s the value to actually align to these design rules}}. As discussed before in Section \ref{sec:case}, Ericsson has developed a family of design rule checkers that automatically check the conformance of the common microservices to the design rules. According to the \ADPA the design rules checkers covered around 50\% of the design rules at the time of conducting the interviews. The \ADPA mentioned that \say{\emph{the goal is to integrate the design rule checkers in CI/CD build that scans all the design rules}}. Such a CI/CD build can provide an instant report of the passed and failed tests. The developers will know which design rules have not been met and find a solution to conform to them efficiently. 
\subsection{Improvement implemented to improve interpretation of design rules and generic product requirements} \label{sec:AutomatingDR} To mitigate the challenge of misinterpreting the design rules, Ericsson implemented the following changes -  
\begin{mybox}{way of working}
\textit{\enquote{\textcolor{black}{Design rules should be written in a way that facilitates auto-testing whenever possible}
}}.
\end{mybox}
\begin{mybox}{conformance}
From \say{\emph{I need to check if I am complaint}} to \say{\emph{The pipelines tell me if I am complaint}}.
\end{mybox}
\begin{mybox}{management and alignment}
From \say{\emph{Rules and compliance as document}} to \say{\emph{Rules and compliance as data}}.
\end{mybox}
In addition, as discussed in Section \ref{IA:MisGPRDR}, the value of implementing a design rule should be clear. One of the criticisms is that there are too many design rules, and implementing them may not be worth it. Creating a design rule should have an intrinsic value and correspond to the main requirements set by the product manager, for example. Ericsson has set a constraint on the number of design rules that can be created: There can never be more than four new design rules every three weeks in the new microservice maturity release.

Another ongoing improvement measure taken by Ericsson is to reduce the cost of implementing the design rules. Therefore, in addition to automating the design rules checking, the implementation of some design rules is also automated.
\subsection{Challenge 3: Difficulty in reusing common microservices} \label{Sec:consumerchallenges}
\begin{figure*}[!t]
    \centering
    \includegraphics[width=\textwidth]{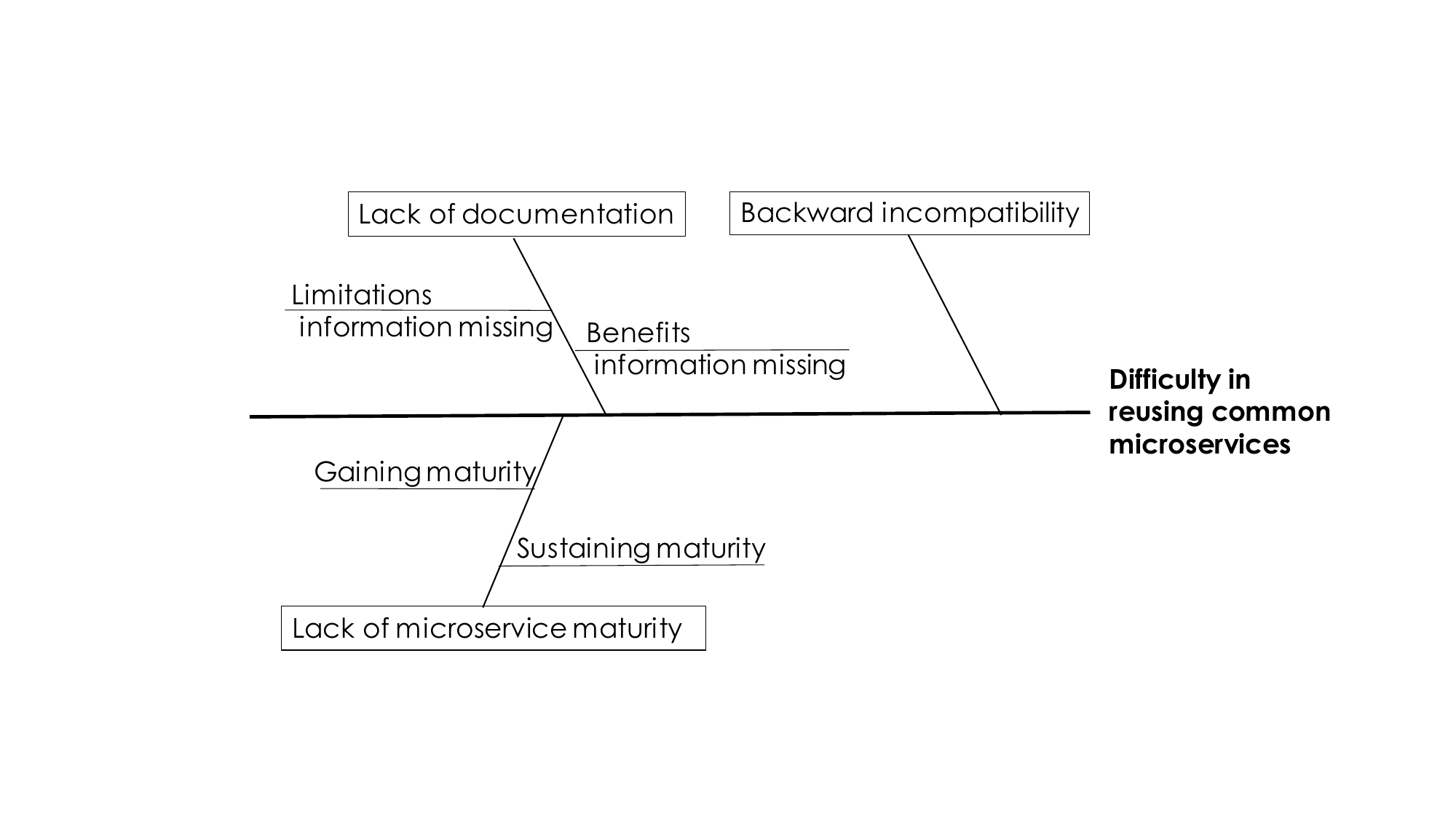}
    \caption{Reasons inhibiting reuse of common microservices.}
    \label{fig:ReusingCommonMicroservices}
\end{figure*}

The consumers mentioned challenges that hinder them from reusing common microservices. The reasons making it difficult to reuse the common microservices are depicted in Figure \ref{fig:ReusingCommonMicroservices} and described in this section. 
\subsubsection{Lack of documentation on microservice purpose, use cases, and limitations} At the time of the interviews, the available documentation for the common microservices in the Marketplace (see Section \ref{Sec:Marketplace}) was limited to deployment and integration. It lacked details on what use cases the microservices implement and their limitations. As mentioned by a \PlatA and \PlatM, \say{\emph{The documentation doesn’t provide any details on limitations... we need to know what it [microservice] can’t do.}}
\subsubsection{Service maturity} According to a \AppCM the new common microservices were relatively less mature and had more issues, such as security vulnerabilities and missing features. In addition, a technical leader mentioned that when already maturity services are changed (e.g., with new features), then \say{\emph{there needs to be some slack for the team in that service to get to the desired quality and maturity level. But it is frustrating that it did not continue.}} Therefore, it takes some time for the common microservices to get mature and continuous effort must be put in to sustain the maturity when new features are added.
\subsubsection{Breaking backward compatibility} The common microservices are updated and released frequently. The consumer teams must also upgrade to the newer microservice versions, but the latest versions sometimes break backward compatibility. As a consequence of backward incompatibility, the applications 
\textcolor{black}{need to synchronize the sequence of upgrading of those multiple microservices that are affected by the upgrade}. A \AppSPM added that \say{\emph{today in existing software, upgrading all the installations can have a huge cost}}. 
\subsection{Improvement suggestion 3: Make common microservices attractive for consumers}

The interviewees mentioned suggestions to make the common microservices easier to use and attractive to consumers. 
\subsubsection{Enhanced common microservices} The common microservices must include some design rules and generic product requirements (GPRs) implementation such as security compliance. According to a \AppSPM \say{\emph{GPRs needs to be driven from a centralized perspective and not from each team contributing.}}  In addition, consumers identify the need to test common microservices rigorously to avoid last-minute blocking issues. The common microservices should also be easy to integrate as stated by  \AppSPM \say{\emph{the services should have a small footprint}}.  
\subsubsection{Ensure upgradeability} The owners of the common microservices teams should consider the feedback from consumers to get a big picture of how the services will be used. For example, one of the \AppCM added that the common microservice owners should know \say{\emph{how the common microservice is used. What are the requirements from the customers' perspective? They need to make sure that it is upgradeable and manageable.}}
\subsubsection{Provide guidelines for integration and configurations} Consumers want more concrete guidelines on how to integrate the common microservices. As seen in Section \ref{Sec:Costs}, integrating common microservices requires additional effort. The lack of guidelines will increase the integration effort. The \AppSPM mentioned \say{\emph{we would need to have much clear rules about how are we supposed to integrate, configure the service and so on.}} 
The \AppCM mentioned that it should be easy to configure the services by changing the HELM charts to expose the parameters that could be configured. 
\subsection{Improvements implemented to make common microservices attractive to consumers}\label{Deprecation} As mentioned in Section \ref{Sec:consumerchallenges}, consumers mentioned three main reasons that made reusing common microservices difficult: Lack of documentation, microservice maturity, and backward incompatibility. The common microservices have reached a high maturity level, and therefore the lack of maturity is no longer a big challenge. The ADP Anchor unit baselines the service maturity staircase (see Figure \ref{fig:Staircases}) with expectations on the services every three weeks, which the microservice teams then implement to keep up with the latest expectations. Automatic design rule checkers (discussed in \ref{sec:AutomatingDR}) help keep up with the latest expectations.
The documentation of the common microservices, i.e., the user guides, is richer now and includes details on integrating and deploying. Therefore, the common microservices documentation and maturity are no longer a challenge. However, the new microservices, particularly those that are contributed by the non-centrally funded teams, may have the same challenges as those produced by the centrally funded teams once had.

To tackle the backward incompatibility challenge, Ericsson already had a deprecation process. If a change is incompatible, then a deprecation is registered. The deprecation will go through an architecture review and a business review before being approved, and the consumers have the opportunity not to approve the deprecation. The deprecation period was for three months. The process worked well. 
\textcolor{black}{However, the consumers complained about the deprecation period being too short}. Ericsson has now updated the deprecation process where the deprecation period is extended to 13 months. It is possible to have a shorter deprecation period with a strong business case, for example, the security hardening or high costs in keeping both old and new solutions active. The longer deprecation period allows the teams to use the common microservices and deploy at the customer end and continue to receive support with fixes for a relatively long period. 
\subsection{Challenge 4: Differences in producer and consumer contexts:} \label{sec:environmentDifferences}
\todo[inline]{Can we collect stats on this from the Ericsson laptop? How many different cloud environments?} The interviewees mentioned challenges related to the differences in the release cycles and the differences in the cloud-native journey between the producers and consumers. We elaborated on the differences in this section. 
\subsubsection{Lifecycle management} The centrally funded teams have a relatively shorter release cycle, which means the newer versions are released much more quickly than the release cycle of the corresponding consumer teams. 

The managers of the consumer teams shared that the centrally funded teams release too frequently. One of the \AppCM mentioned that \say{\emph{They [centrally funded teams] release every month or three weeks and we release, for example, twice a year.}} \textcolor{black}{Frequent releases are needed to ensure that the releases are not too large and to promptly respond to the several feature requests and bug reports the centrally funded teams receive. The consumer teams expect a fast response, as mentioned in Section \ref{Sec:turnaroundtime}; they do not want to wait too long. Therefore, frequent releases are needed; however, this may also result in a challenge for some consumer applications}. 
\subsubsection{Deployment} As the consumers of common microservices are growing within Ericsson, it is becoming a challenge to ensure technical alignment of the interfaces. 
When multiple applications with different cloud-native maturity are packaged together in one offering, it becomes challenging to have alignment in the deployment and configurations. One of the \ADPCA mentioned that \say{\emph{if one application has gone very far and someone just started[cloud native journey], you would have big problems.}} In addition, a \AppSPM mentioned that different applications might integrate the microservices differently, reducing the reuse benefits such as having the same look and feel of the common microservices.

The challenges due to the differences in producer and consumer contexts will be mitigated in the long term as the common microservices become more mature and the teams become fully cloud-native. 
\subsubsection{Alignment to the purpose and scope of microservices}
The interviewees reported coordination challenges between the consumers and producers. 
Any fixes or new requirements consumers request should align with the purpose and scope of the microservices. Conflicts may arise when consumer requests do not align with the common microservice purpose and scope. In addition, some teams may not want a requirement to be part of the microservice that some other teams may want. The alignment of the purpose and scope involves a lot of coordination, which according to a \AppTL can be \say{\emph{very time-consuming, and there can be long debates}}. Once the feature requests are approved, they are prioritized and added to the roadmap: indicating the features planned to be added along with the timeline. When multiple teams request a requirement, it may be easier to prioritize the request, \textcolor{black}{justifying the business need}. However, \textcolor{black}{smaller teams may be affected} according to a \AppCM \say{\emph{different applications have different weights..... we, as a small product, have less influence and weight on their[producers] roadmap and backlog.}} 


\subsection{Improvement implemented to address the differences in producer and consumer contexts} One of the differences in context is the difference in the lifecycle of the common microservices and the consumer teams. A shorter cycle of the common microservices is needed to keep up with the feature requests and bug reports. In addition, extending the deprecation period (discussed in Section \ref{Deprecation}) helps to mitigate the challenge of frequent updates. 

In addition, the interviewees perceived the alignment of the deployment to be challenging. Ericsson has now defined different phases of flexible deployment. The definition of the deployment phases allows several teams, in runtime, to instantiate so they can share instances of the common microservices. 
The teams should use the common microservices in an aligned way. The guidelines for integration and configuration are more prescriptive.

\section{Discussion}
\label{sec:discussion}

\todo[inline] {
Possible themes: Microservice as a reusable unit, CI/CD as a support for large-scale reuse (one reference about Hybrid CI/CD?- Ali, Nazakat, Horn Daneth, and Jang‐Eui Hong. "A hybrid DevOps process supporting software reuse: A pilot project." Journal of Software: Evolution and Process 32.7 (2020): e2248.), InnerSource.
}


The large-scale reuse of common microservices at Ericsson has been made possible with the help of a combination of aspects including InnerSource practices, a cloud-native microservices architecture together with DevOps practices. In the following sub-sections, we further discuss the results of our study from these aspects. In addition, we will also discuss these results in relation to the existing relevant works.

\subsection{InnerSource and reuse}

Ericsson, being a large organization, has many potential consumers of common microservices - which implies that when they reuse and integrate the common microservices into their applications, they will not only report bug fixes but will also ask for changes and new features. As practitioners in the ADP program put it - \textit{\enquote{the success of such microservices may become their failure}} - i.e., it may become impossible for the teams of such common microservices to handle a large number of change requests and new requirements by the consumers of these services. Such a scenario led to a bottleneck phenomenon wherein the teams owning the common microservices found it hard to process their backlogs. This bottleneck phenomenon has also been reported previously (see for example \cite{riehle2016inner,cooper2018adopting}) wherein the organizations attempted to develop and sustain a platform-based development. In such a development approach, a centrally funded platform unit is responsible for maintaining the platform without many contributions from the other organizational units the use the platform - resulting in a silo structure \cite{riehle2016inner}. The ADP program was aware of the potential pitfalls of such an approach - i.e., a platform way of looking at and working with the ADP ecosystem. They actively worked to avoid and address this bottleneck phenomenon with the help of different initiatives, which are highlighted in the following sections.

\subsubsection{InnerSource roles, practices and guidelines}
ADP Program and its Anchor unit is responsible for developing policies, mechanisms and guidance for sustaining the ecosystem. ADP Anchor unit takes lead in not only developing and updating the guidelines and policies for InnerSource contributions but it also on boards and mentors new contributors. Such a unit is necessary for sustaining the InnerSource initiatives in organizations \cite{riehle2016inner}. The participants in our study recognized and appreciated the role the ADP Anchor unit is playing in continuously updating the guidance on InnerSource. The introduction and definition of roles such as \enquote{\textit{Guardian}} and \enquote{\textit{Trusted Committer}} - based on the recommendations from InnerSource Commons \cite{ISCommons_IP} - has helped in clarifying the roles and responsibilities. 

Due to the focus on their own work assignments, developers may not have any bandwidth to contribute to a common asset - which is one of the main reasons why it is hard to attract InnerSource contributions \cite{morgan2011exploring}. In the studied case as well, it took a while before the contributions to the common microservices started to increase. The sustained efforts by the ADP Program and its Anchor Unit, and management support have started to show positive results in terms of more InnerSource contributions. 
\begin{mybox2}{InnerSource and Reuse}

a) In large organizations like Ericsson, the central unit like ADP Anchor, is critical to take the InnerSource initiative forward. The stakeholders - e.g., potential consumers and contributors to the common microservices - need guidance, training, and a clear description of their responsibilities (e.g., what will happen after they have contributed to a service).\\
b) The InnerSource policies and guidance need to be continuously updated in response to the current challenges being faced by different stakeholders.\\
c) Lack of funding and available time are among the main challenges that inhibit consumers from prioritizing InnerSource contributions. Ensuring sustained top management support is critical to resolving the funding related concerns of the middle managers.

\end{mybox2}

\subsubsection{Marketplace and reuse}

One of the challenges of practicing reuse is the lack of discoverability of the reusable assets \cite{barros2019exploratory,bauer2014exploratory,chen2022reuse}. Often the companies make their reusable assets available in a shared repository. However, such repositories may not have the best search functions and filters \cite{chen2022reuse}. In addition, in the large organization, they may not be accessible and visible to all teams and units. The Marketplace of Ericsson not only allows all potential consumers of the common microservices to efficiently search and filter their desired services, but it also facilitates InnerSource contributions. It is a one-stop place where consumers and contributors can not only explore the relevant microservices but can also find their corresponding documentation and links for integrating and contributing to the services. The study participants appreciated the role Marketplace is playing in the development, discovery, integration and InnerSource contributions to the common microservices. Similar shared spaces have also been established by other companies as well, such as by Lucent Technologies \cite{gurbani2006case}, Philips \cite{lindman2010open}, Nokia \cite{lindman2010open}, Hewlett-Packard \cite{melian2008lost}, IBM \cite{vitharana2010impact}, and Zalando, Philips Healthcare and PayPal \cite{morgan2021share}. Like Marketplace, these shared spaces are normally web-based portals available to everyone within the organization where it is possible to search the common assets and access their documentation and links to code repositories, etc. The concept of Maturity and Reusability staircases being used in the Ericsson's Marketplace offers a unique way of classifying the common microservices, which we did not find in the reported cases about the similar shared spaces.

The service maturity is helpful for producers as well as consumers of the microservices. The maturity levels provide a roadmap to the microservices' teams (producers), clarifying which design rules they need to implement to move to the higher level. Likewise, the consumers can not only search and filter mature microservices, but also become familiar with what can be expected from a microservice based on its maturity level. The reusability staircase, on the other hand, is helpful in identifying those microservices that have (not) been successfully reused by one or more applications. The study participants recognized that these two staircases help them in planning their work - producers in planning the roadmap of their microsevices and consumers in identifying the relevant microservice for reuse.

\begin{mybox2}{Marketplace and Reuse}

a) The development with and for reuse at a large scale is not possible without putting in place an effective portal - like Marketplace - where it is possible to easily find and filter reusable assets.\\
b) The consumers of the reusable assets also need integration and configuration support. Therefore, such shared spaces also need to provide support and instructions to the consumers on how to integrate the reusable assets into their applications.\\
c) Such a shared space is also crucial for the InnerSource initiative - for example by providing guidance (e.g., contribution guidelines), links (e.g., to code repos) and other support (e.g., tutorials) to the potential contributors.
\end{mybox2}


\subsection{Microservices architecture, DevOps and reuse}

The ADP ecosystem uses a cloud-native microservices-based architecture with containers. In the reported case, the reusable assets are the microservices that implement the common functions required by different applications across Ericsson. Software reuse has been around for decades - but the microservices are a relatively recent phenomenon. Several large companies (e.g., Netflix, Microsoft) are using microservices to deliver their services and products. In the studied case, the ADP program and the involved practitioners found microservices as a good unit for reuse due to their size and independent deployment, which is in line with a few reported experiences of other companies - for example - Tizzei et al. \cite{tizzei2017using},  Assun{\c{c}}{\~a}o \cite{assunccao2021multi}, and Gouigoux and Tamzalit \cite{gouigoux2017monolith} also noted that the use of microservices has helped in improving the software reuse. However, these studies do not report the microservices' reuse at such a large scale, which is the focus of our study. Our study offers a unique perspective of practicing microservices' reuse at such a large scale and that too in combination with contemporary software engineering practices - InnerSource and DevOps practices in a cloud-native context. There is a need to conduct more industrial case studies to further characterize and understand such contemporary reuse - i.e., practicing reuse of modern reusable assets (i.e., microservices) with the help of new practices (e.g., InnerSource and DevOps) and technology (e.g., Dockers, Kubernetes).

The large scale reuse of microservices in Ericsson has two dimensions: a) a large number of common microservices (over 280 at the time of the study) that can be potentially reused by any application, and 2) a large number of applications in Ericsson that reuse the common microservices (e.g., some common microservices are used in over 40 applications). The combination of these two dimensions implies that it becomes critical to develop a mechanism wherein new versions of common microservices are made available for integration by the applications that consume these services. Ericsson's CI/CD strategy provides such a mechanism wherein the microservices own CI pipelines are connected with the consumer applications' assembly pipelines with the help of Spinnaker pipelines. This mechanism ensures that when a new version of a common microservice is released, it becomes available to the consumer applications staging environment for their testing process. According to the study participants, this mechanism is key to sustain the large scale development and reuse of microservices across Ericsson. The study findings indicate a relationship between the DevOps (i.e., CI/CD practices) and large scale reuse of microservices. However, we have not found many studies that investigate the potential relationship of DevOps and reuse. In addition, the concept of design rules and their automated checking is playing a crucial role in the automated testing and continuous delivery of microservices. The DR checkers are much appreciated by the microservices teams as they allow them to automatically check the compliance of their microservices with the ground principles, constraints, and requirements.

\begin{mybox2} {Microservices and Reuse}

a) Due to their size and focus on implementing specific functionality, microservices are potentially a good unit for reuse - especially when implemented in a cloud-native context with containers to support flexible and independent deployment. Companies can improve their reuse potential by moving from a monolith to microservices architecture.\\
b) The large scale reuse is only possible if organizations put in place effective delivery and deployment mechanisms. In absence of appropriate CI/CD strategy and technology, it won't be possible for the applications to efficiently reuse common microservices.\\
c) Establishing and ensuring compliance to common principles and rules is crucial in such large scale initiatives to avoid conflicting implementations. However, manually testing the compliance to a large number of rules will become a bottleneck in itself. Automation - like Ericsson's DR checkers - provides an efficient mechanism to developers for automatically testing the compliance of their microservices.
\end{mybox2}


\section{Conclusion}
\label{sec:conclusion}
The widespread reuse of microservices was possible in Ericsson due to the adoption of contemporary reuse practices such as InnerSource and DevOps. Adoption of contemporary reuse inevitably incurs costs such as the need to update competence to learn contemporary practices. However, the costs result in long-term benefits such as improved quality, productivity, customer experience, and way of working. Moreover, the adoption of these practices was not without challenges - it rather is a journey in which several stakeholders have to continuously adapt their way of working to address the challenges they face while adopting the new ways of working. At the beginning of the journey, both reusable asset producers (e.g., increased workload created a bottleneck) and consumers (e.g, lack of maturity of the reusable asset) faced challenges. Ericsson mitigated the challenges through the adoption of InnerSource, automating the implementation and checking of design rules and general requirements, and enhancing the maturity of common microservices. 

Ericsson found the InnerSource contributions made to the common microservices as the key enabler to the growth and sustainability of microservice reuse. As part of our future work, we intend to investigate InnerSource contributions in depth to identify factors leading to making successful InnerSource contributions.  




\begin{acknowledgements}
This work was supported by the Knowledge Foundation through the OSIR project (reference number 20190081) at the Blekinge Institute of Technology, Sweden. 
\end{acknowledgements}

\section*{Conflict of Interest} 
The authors have no conflicting interests to declare concerning the content of this manuscript.

\section*{Data Availability Statements}
We collected data through interviews, group discussions, reuse-related portals, and company documents. In Sections \ref{sec:resultsRQ1} and \ref{sec:challenges}, we provide interview quotes to support our findings. The raw interview transcripts cannot be made publicly available to ensure the anonymity of the interviewees and protect details that allow identification of the interviewee or the project/team. The data extracted from the company documents such as Figures \ref{fig:Staircases} are provided in the manuscript itself with permission from Ericsson. Figure \ref{fig:ContTrend} is generated from the reuse-related portals. The raw data from the reuse-related portals cannot be made available as it includes company sensitive data. No datasets were generated or analyzed for this study. 



%
\bibliographystyle{plain}
\bibliography{References.bib}

\end{document}